\documentclass[a4paper,fleqn,usenatbib]{mnras}

\usepackage{newtxtext,newtxmath}
\usepackage[utf8]{inputenc} 
\usepackage{threeparttable}
\usepackage{lscape}
\usepackage{upgreek}
\usepackage{amssymb}
\usepackage[T1]{fontenc}
\usepackage{ae,aecompl}
\usepackage[figuresright]{rotating}
\usepackage{float} 

\usepackage{graphicx}	% Including figure files
\usepackage{amsmath}	% Advanced maths commands
\usepackage{amssymb}	% Extra maths symbols

\newcommand{\clk}{\textrm{clk}}

\newcommand{\poster}{P_{\textrm{pos}}}
\newcommand{\prior}{P_{\textrm{pr}}}
\newcommand{\likeli}{\Lambda}

\newcommand{\mn}{\textsc{Multinest}}
\newcommand{\ftwo}{\textsc{fortytwo}}

\newcommand{\trs}{\mathbfss{T}}
\newcommand{\cov}{\mathbfss{C}}
\newcommand{\covdash}{\mathbfss{C}'}
\newcommand{\desm}{\mathbfss{D}}

\newcommand{\tres}{\boldsymbol{t}}

\title[A pulsar-based timescale]{ A pulsar-based timescale
 from the International Pulsar Timing Array}

\author[G. Hobbs et al.]{
G. Hobbs,$^{1}$\thanks{E-mail: george.hobbs@csiro.au}
L. Guo,$^{2}$
R.~N. Caballero,$^{3,4}$
W. Coles,$^{5}$
K. J. Lee,$^{4}$
R. N. Manchester,$^{1}$
\newauthor
D. J. Reardon,$^{6}$
{D. Matsakis},$^{7}$
M. L. Tong,$^{8}$
{Z. Arzoumanian},$^{9}$
{M. Bailes},$^{6}$
\newauthor
{C. G. Bassa},$^{10}$
{N. D. R. Bhat},$^{11}$
{A. Brazier},$^{12, 13}$
{S. Burke-Spolaor},$^{14,15}$
{D. J. Champion},$^3$
\newauthor
{S. Chatterjee},$^{12}$
I. Cognard,$^{16,17}$
{S. Dai},$^{1}$
{G. Desvignes},$^3$
{T. Dolch},$^{18}$
{R.~D.~Ferdman},$^{19}$
\newauthor
{E. Graikou},$^3$
{L. Guillemot},$^{16,17}$
{G. H. Janssen},$^{10,20}$
{M. J. Keith},$^{21}$
{M. Kerr},$^{22}$
\newauthor
{M. Kramer},$^3$
{M. T. Lam},$^{14}$
{K. Liu},$^3$
{A. Lyne},$^{21}$
{T. J. W. Lazio},$^{27}$
R. Lynch,$^{24,14}$
\newauthor
{J. W. McKee},$^3$
{M. A. McLaughlin},$^{14,15}$
{C. M. F. Mingarelli},$^{25}$
\newauthor
{D. J. Nice},$^{26}$
{S. Os{\l}owski},$^6$
{T. T. Pennucci},$^{27}$
{B. B. P. Perera},$^{21}$
{D. Perrodin},$^{28}$
\newauthor
{A. Possenti},$^{28,29}$
{C. J. Russell},$^{30}$
{S. Sanidas},$^{21}$
{A. Sesana},$^{31}$
{G. Shaifullah},$^{10}$
\newauthor
{R. M. Shannon},$^{6,32}$
{J. Simon},$^{23}$
{R. Spiewak},$^{6}$
{I. H. Stairs},$^{33}$
{B. W. Stappers},$^{21}$
\newauthor
{J. K. Swiggum},$^{34}$
{S. R. Taylor},$^{35}$
{G. Theureau},$^{16,17,36}$
{L. Toomey},$^{1}$
{R. van Haasteren},$^{23}$
\newauthor
{J. B. Wang},$^{37}$
{Y. Wang},$^{38}$
{X. J. Zhu}$^{39}$
\\
\\
Affiliations given at the end of the paper 
}

\date{Accepted XXX. Received YYY; in original form ZZZ}

\pubyear{2016}

\begin{document}
\label{firstpage}
\pagerange{\pageref{firstpage}--\pageref{lastpage}}
\maketitle

\begin{abstract}
We have constructed a new timescale, TT(IPTA16), based on observations of radio pulsars presented in the first data release from the International Pulsar Timing Array (IPTA). We used two analysis techniques with independent estimates of the noise models for the pulsar observations and different algorithms for obtaining the pulsar timescale. The two analyses agree within the estimated uncertainties and both agree with TT(BIPM17), a post-corrected timescale produced by the Bureau International des Poids et Mesures (BIPM).
We show that both methods could detect significant errors in TT(BIPM17) if they were present. We estimate the stability of the atomic clocks from which TT(BIPM17) is derived using observations of four rubidium fountain clocks at the US Naval Observatory. Comparing the power spectrum of TT(IPTA16) with that of these fountain clocks suggests that pulsar-based timescales are unlikely to contribute to the stability of the best timescales over the next decade, but they will remain a valuable independent check on atomic timescales. We also find that the stability of the pulsar-based timescale is likely to be limited by our knowledge of solar-system dynamics, and that errors in TT(BIPM17) will not be a limiting factor for the primary goal of the IPTA, which is to search for the signatures of nano-Hertz gravitational waves.
~\\

\end{abstract}


%%%%%%%%%%%%%%%%%%%%%%%%%%%%%%%%%%%%%%%%%%%%%%%%%%

%%%%%%%%%%%%%%%%% BODY OF PAPER %%%%%%%%%%%%%%%%%%

\section{Introduction}

International Atomic Time (TAI) is the basis for terrestrial time.  It is also used to form Coordinated Universal Time (UTC), which approximates the rotational phase of the Earth and provides a realisation of the theoretical timescale Terrestrial Time (TT) through ${\rm TT(TAI)} = {\rm TAI} + 32.184$\,s.  Once TAI is defined, it is not changed, but it is reviewed annually and departures of TAI from the SI second are incorporated into a post-processed timescale published by the Bureau International des Poids et Mesures (BIPM; \citealt{pet03b}) as TT(BIPMXY), where XY indicates the year of creation. The version we have used for this analysis is TT(BIPM17), which is available from \url{ftp://ftp2.bipm.org/pub/tai/ttbipm/TTBIPM.17}. 

Atomic timescales continue to improve in accuracy and stability \citep{pet13}. Such timescales are created from an ensemble of atomic clocks (e.g. hydrogen masers) which have good short term stability but are weaker over longer periods. To provide stability over years and decades TAI is `steered' by comparison with `primary and secondary representations of the second' (presently caesium and rubidium fountains respectively) to keep the TAI second as close as possible to the SI second \citep{app11}. This leads to a timescale which is statistically non-stationary and for which the stability over decades is difficult to determine. It would be valuable to have an independent timescale with comparable stability over decades.

\citet{gp91}, \citet{mf96}, \citet{pt96}, \citet{rod11}, \citet{hcm+12}, \citet{mghc17} and \citet{ygz17}  have shown that a timescale based on the spin of radio pulsars with millisecond periods (MSPs) can have a stability comparable to that of atomic timescales.  Whereas  normal (longer period) pulsars often show irregular rotation, such as glitch events or low-frequency timing noise, MSPs are significantly more stable. Pulse times of arrival (ToAs) from MSPs can also be determined more precisely than those from normal pulsars because they are shorter and there are more pulses to average.  

More than 50 MSPs are currently being observed as part of the International Pulsar Timing Array (IPTA) project, with the primary goal of detecting ultra-low-frequency gravitational waves \citep[e.g.,][]{haa+10,vlh+16}.  The IPTA  combines the Parkes PTA (PPTA; \citealt{mhb+13}) in Australia with the European PTA (EPTA; \citealt{dcl+16}) and with the North American PTA (NANOGrav; \citealt{abb+18}, \citealt{abb+15}) and it will continue to expand in the future.

The IPTA team expects to identify the quadrupolar signature of gravitational waves unambiguously by searching for the  correlations between the timing residuals of different pulsars (see, e.g., \citealp{vlh+16}).  As shown by \citet{thk+16} and references therein, other processes can also produce correlated timing residuals.  Errors in the planetary ephemeris will introduce an approximately dipolar correlation between pulsars, and timescale errors will introduce a monopolar correlation. Since the strongest gravitational waves are expected to have long periods, the stability of the reference timescale over decades is important for gravitational wave searches.

Two methods have been used for extracting a pulsar-based time standard from PTA data, and for comparing it with a given realisation of TT.  The first, described in \citet{hcm+12}, uses a frequentist-based method that includes the clock signal as part of the pulsar timing model and carries out a global, least-squares-fit to estimate that signal. The second, a Bayesian technique described in \citet{cll+16}, uses a maximum-likelihood estimator and optimal filtering, described in \citet{lbj+14}, to estimate the clock signal.

In this paper, we:
\begin{itemize}
\item make use of the first IPTA data set to produce a pulsar-based time standard TT(IPTA16),
\item improve the \citet{hcm+12} algorithm by accounting for non-stationarity in the noise processes (as described by \citealt{rhc+16}),
\item extend the \citet{cll+16} algorithm to estimate both the covariance of the noise processes and the clock signal using independent Bayesian algorithms. This improves the accuracy of the uncertainty of clock signal waveform,
\item compare the two different methods for developing the pulsar-based time standard, and
\item provide a direct comparisons between TT(IPTA16) with respect to TT(BIPM17), and the best atomic frequency standards at the US Naval Observatory.
\end{itemize}

\section{The Data Set and Analysis}\label{sec:data}

The pulsar timing analysis is a well-developed process of fitting a model, which describes both the pulsar and the propagation of the pulses to Earth, to a set of observed pulse ToAs (see \citealt{hem06} for details of how this process is implemented within the \textsc{tempo2} software package). For this work, we use ToAs from the first IPTA data release by \citet{vlh+16} and \citet{lsc+16}.   The data set consists of observations of 49 pulsars from the EPTA telescopes (Lovell, Effelsberg, Westerbork and Nan\c cay), the NANOGrav telescopes (Arecibo and Green Bank) and the Parkes telescope. Of the 49 pulsars\footnote{The data set for one of these pulsars, PSR~J1939$+$2134, was removed from our sample.  The data are affected by significant low-frequency timing noise and the pulsar does not contribute to determination of the clock signal.}, 13 are solitary and 36 are in binary systems\footnote{This includes PSR~J1024$-$0719, which is thought to be in a long-period orbit; see \citet{bjs+16} and \citet{kkn+16}.}. The pulsars have been observed in numerous observing bands from around 300 to 3000\,MHz.  The observations are performed with an irregular observing cadence of 2 to 4 weeks (depending on the pulsar and the telescope). For the basic timing analysis we used {\sc tempo2} \citep{hem06} with the DE421 solar-system ephemeris \citep{fwb09}.

The ToAs were estimated by fitting a pulse template to the observed average pulse profile. This provides both a ToA and an estimate of its measurement error. The uncertainties of the observed ToAs are more complex than the measurement errors (see \citealt{vs18} for a review). For instance, the strength of the observed pulse varies rapidly because of interstellar scintillation (see, e.g., \citealt{ric70}), and so the signal-to-noise ratio changes rapidly. The pulse shape itself shows significant jitter from pulse to pulse (see, e.g., \citealt{lkl+12,sod+14,lma+18}). Such effects cause independent errors in each measurement, which we refer to as ``white noise". 

The timing residuals are also affected by low frequency noise processes, which we refer to as ``red noise". The pulsar timing method (details are provided in \citealt{ehm06}) relies on an initial set of models that includes the pulsar spin frequency and its spin-down rate, its position on the sky and proper
motion and parallax, the binary orbital parameters if any, and the electron column density of the interstellar medium (known as the dispersion measure, DM). It also includes instrumental parameters such as phase differences between receivers.  However, pulsars often spin irregularly, the interstellar medium changes, and the observing hardware and software change (see details in e.g., \citealt{mhb+13,lsc+16,abb+18}). Such effects lead to red noise in the timing residuals. The covariance matrices of all sources of noise must be estimated to optimise the least-squares (or maximum-likelihood) fits and to obtain valid uncertainties on the parameters of the timing model.   Recent reviews of such issues in pulsar timing residuals in relation to PTA experiments were recently published by \citet{hd17}  and \citet{tib18}.

\citet{vlh+16} provided the  first IPTA data release in three different forms: (A) a ``raw" format with only minimal pre-processing, (B) a format usable for standard \textsc{tempo2} analysis and (C) a data set using the Bayesian parameterisation for \textsc{temponest} analysis.  These data sets are available from \url{http://www.ipta4gw.org}.  The primary difference between these three data sets is the way in which the noise models are included. For our analysis we started with the IPTA data combination `A', which is the raw form without any noise models.

The derived pulsar timescale described in this paper is named TT(IPTA16) as our data set is based on \citet{vlh+16}, which was published in 2016. However, we note that the data set only extends until the end of 2012.  

\subsection{Data and processing for the frequentist analysis}

\begin{table*}
\caption{Summary of the data used.}\label{tb:summaryTable}
\begin{tabular}{p{1.5cm}lp{0.4cm}llp{2.0cm}p{1.9cm}p{1.7cm}p{1.7cm}}
\hline
Pulsar & MJD range & Span & Frequency & Ntoa & Telescopes & Max diff. (ns) & Ratio & Bayesian  \\
 & & & range & & & (Significance) & (Rank) & rank \\
  &  & (yr) & (MHz) & & &  \\
\hline
J0030+0451 & 53333--55925  & 7.1  & 1346--2628  & 723  & A,E,N & 11 (\#23) & 1.00 (\#25) & \#9\\
J0034$-$0534 & 53670--55808  & 5.9  & 1394--1410  & 57  & N & 1 (\#48) & 1.00 (\#26) & \#10\\
J0218+4232 & 50371--55925  & 15.2  & 1357--2048  & 757  & E,J,N,W & {\bf 8 (\#27)} & 0.98 (\#42) & -\\
J0437$-$4715 & 50191--55619  & 14.9  & 1295--3117  & 4320  & P & 254 (\#3) & 1.54 (\#3) & Top\\
J0610$-$2100 & 54270--55926  & 4.5  & 1366--1630  & 347  & J,N & {\bf 5 (\#29)} & 0.99 (\#38) & \#20\\
\\
J0613$-$0200 & 50932--55927  & 13.7  & 1327--3100  & 1460  & E,G,J,N,P,W & {\bf 24 (\#17)} & 0.93 (\#46) & -\\
J0621+1002 & 50693--55922  & 14.3  & 1354--2636  & 568  & E,J,N,W & {\bf 3 (\#39)} & 0.99 (\#40) & -\\
J0711$-$6830 & 49589--55619  & 16.5  & 1285--3102  & 427  & P & 74 (\#7) & 1.15 (\#6) & -\\
J0751+1807 & 50363--55922  & 15.2  & 1353--2695  & 1124  & E,J,N,W & 34 (\#10) & 1.08 (\#10) & -\\
J0900$-$3144 & 54285--55922  & 4.5  & 1366--2206  & 575  & J,N & 4 (\#31) & 1.00 (\#29) & \#15\\
\\
J1012+5307 & 50647--55924  & 14.4  & 1344--2636  & 1117  & E,G,J,N,W & {\bf 44 (\#8)} & 0.79 (\#48) & -\\
J1022+1001 & 50361--55923  & 15.2  & 1341--3102  & 1133  & E,J,N,P,W & 31 (\#11) & 1.25 (\#4) & -\\
J1024$-$0719 & 50118--55922  & 15.9  & 1285--3102  & 804  & E,J,N,P,W & 11 (\#25) & 1.00 (\#31) & -\\
J1045$-$4509 & 49406--55620  & 17.0  & 1260--3102  & 510  & P & {\bf 14 (\#22)} & 0.96 (\#44) & -\\
J1455$-$3330 & 53375--55926  & 7.0  & 1246--1699  & 395  & J,N & {\bf 3 (\#34)} & 0.99 (\#37) & \#6\\
\\
J1600$-$3053 & 52302--55919  & 9.9  & 1341--3104  & 999  & G,J,N,P & 28 (\#15) & 1.01 (\#17) & \#8\\
J1603$-$7202 & 50026--55619  & 15.3  & 1285--3102  & 381  & P & {\bf 18 (\#19)} & 0.97 (\#43) & -\\
J1640+2224 & 50459--55924  & 15.0  & 1353--2636  & 532  & A,E,J,N,W & 27 (\#16) & 1.09 (\#9) & \#13\\
J1643$-$1224 & 49422--55919  & 17.8  & 1280--3102  & 1066  & E,G,J,N,P,W & 23 (\#18) & 1.02 (\#15) & -\\
J1713+0747 & 49422--55926  & 17.8  & 1231--3102  & 1833  & A,E,G,J,N,P,W & 1319 (\#1) & 3.12 (\#1) & Top\\
\\
J1721$-$2457 & 52076--55854  & 10.3  & 1360--1412  & 152  & N,W & 2 (\#45) & 1.00 (\#32) & -\\
J1730$-$2304 & 49422--55921  & 17.8  & 1285--3102  & 480  & E,J,N,P & 37 (\#9) & 1.10 (\#8) & -\\
J1732$-$5049 & 52647--55582  & 8.0  & 1341--3102  & 190  & P & 10 (\#26) & 1.04 (\#12) & -\\
J1738+0333 & 54103--55906  & 4.9  & 1366--1628  & 206  & N & {\bf 3 (\#35)} & 0.99 (\#39) & \#17\\
J1744$-$1134 & 49921--55925  & 16.4  & 1264--3102  & 823  & E,J,N,P & 260 (\#2) & 2.39 (\#2) & \#4\\
\\
J1751$-$2857 & 53746--55837  & 5.7  & 1398--1411  & 78  & N & 3 (\#40) & 1.00 (\#28) & \#21\\
J1801$-$1417 & 54184--55921  & 4.8  & 1396--1698  & 86  & J,N & 3 (\#38) & 1.00 (\#19) & -\\
J1802$-$2124 & 54188--55903  & 4.7  & 1366--2048  & 433  & J,N & 2 (\#44) & 1.00 (\#24) & \#14\\
J1804$-$2717 & 53747--55915  & 5.9  & 1397--1520  & 76  & J,N & 2 (\#43) & 1.00 (\#20) & -\\
J1824$-$2452A & 53519--55619  & 5.7  & 1341--3100  & 234  & P & 4 (\#33) & 1.00 (\#23) & -\\
\\
J1843$-$1113 & 53156--55924  & 7.6  & 1374--1630  & 174  & J,N,W & {\bf 5 (\#30)} & 0.99 (\#36) & -\\
J1853+1303 & 53371--55923  & 7.0  & 1370--1520  & 102  & A,J,N & 11 (\#24) & 1.01 (\#16) & \#11\\
J1857+0943 & 50459--55916  & 14.9  & 1341--3102  & 625  & A,E,J,N,P,W & 29 (\#12) & 1.10 (\#7) & -\\
J1909$-$3744 & 52618--55914  & 9.0  & 1341--3256  & 1398  & G,N,P & 172 (\#5) & 1.24 (\#5) & Top \\
J1910+1256 & 53371--55887  & 6.9  & 1366--2378  & 106  & A,J,N & 5 (\#28) & 1.00 (\#21) & \#12\\
\\
J1911$-$1114 & 53815--55881  & 5.7  & 1398--1520  & 81  & J,N & 4 (\#32) & 1.00 (\#18) & -\\
J1911+1347 & 54093--55869  & 4.9  & 1366--1408  & 45  & N & 17 (\#20) & 1.03 (\#14) & -\\
J1918$-$0642 & 52095--55915  & 10.5  & 1372--1520  & 265  & G,J,N,W & 29 (\#13) & 1.08 (\#11) & \#5\\
J1955+2908 & 53798--55919  & 5.8  & 1386--1520  & 132  & A,J,N & 2 (\#41) & 1.00 (\#33) & \#16\\
J2010$-$1323 & 54087--55918  & 5.0  & 1366--2048  & 296  & J,N & 17 (\#21) & 1.03 (\#13) & \#18\\
\\
J2019+2425 & 53446--55921  & 6.8  & 1366--1520  & 80  & J,N & 2 (\#47) & 1.00 (\#27) & -\\
J2033+1734 & 53894--55918  & 5.5  & 1368--1520  & 130  & J,N & 2 (\#46) & 1.00 (\#30) & -\\
J2124$-$3358 & 49490--55925  & 17.6  & 1260--3102  & 1028  & J,N,P & {\bf 97 (\#6)} & 0.98 (\#41) & -\\
J2129$-$5721 & 49987--55618  & 15.4  & 1252--3102  & 343  & P & {\bf 29 (\#14)} & 0.96 (\#45) & -\\
J2145$-$0750 & 49518--55923  & 17.5  & 1285--3142  & 1476  & E,J,N,P,W & {\bf 198 (\#4)} & 0.92 (\#47) & -\\
\\
J2229+2643 & 53790--55921  & 5.8  & 1355--2638  & 234  & E,J,N & 3 (\#36) & 1.00 (\#22) & \#19\\
J2317+1439 & 50459--55918  & 14.9  & 1353--2638  & 409  & E,J,N,W & 3 (\#37) & 1.00 (\#34) & \#7\\
J2322+2057 & 53916--55921  & 5.5  & 1395--1698  & 199  & J,N & 2 (\#42) & 1.00 (\#35) & -\\
\hline
\end{tabular}
~\\
Telescope codes: (A) Arecibo, (E) Effelsberg, (G) Green Bank, (J) Jodrell Bank, (N) Nancay, (P) Parkes and (W) Westerbork
\end{table*}

\begin{figure*}
\includegraphics[width=13cm,angle=-90]{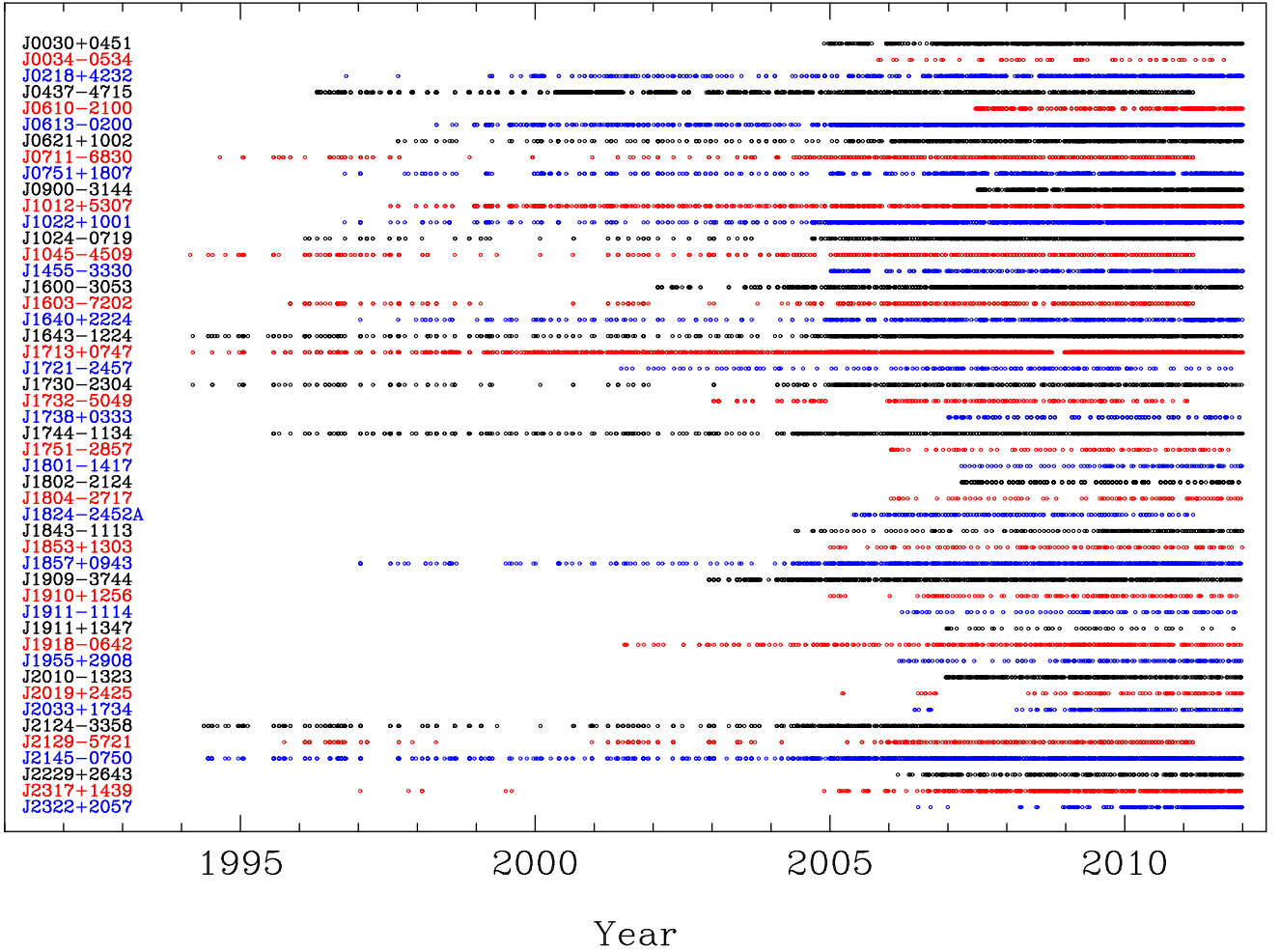}
\caption{Data spans and sampling for the IPTA pulsars used in our analysis. Each point indicates an observation,  but neither the ToA nor its uncertainty are shown. The uncertainties typically are smaller for more recent data. }\label{fg:sampling}
\end{figure*}

We have independently processed the data using both frequentist and Bayesian methodologies.  Here we describe the frequentist processing. The Bayesian analysis is presented in Section~\ref{sec:bayesian}.

For our data set, any irregularities in TT(BIPM17) are much smaller than the noise in the ToAs for any particular pulsar\footnote{This is clear from the results of \citet{hcm+12}, which covered approximately the same data span as the IPTA data processed here.}.  Consequently, we made noise models from the residuals formed from ToAs with respect to TT(BIPM17) and used those models in all fits. This allowed us to avoid iteration of the global clock fit.  Modeling of the noise is perhaps the most difficult part of the timing analysis. The fundamental reason for this is that all pulsars are different and we only have one realization of each pulsar. Thus it is difficult to estimate the covariance matrix without making some constraining assumptions, for example, assuming that the red timing noise is wide-sense stationary and parameterizing its power spectrum with a simple analytical model.  

The noise models refer to the residuals after all the deterministic effects have been removed. Variations in the reference clock can be included in the timing model, estimated and removed from the residuals. So in our frequentist approach to fitting the timing model, we estimate the noise models iteratively. Fortunately such iterations converge quite quickly because least squares solutions are not strongly sensitive to small changes in the noise models.

In Table~\ref{tb:summaryTable} we summarise the data set, providing, in column order, the pulsar J2000 name, the MJD range and corresponding span in years, the range of the observing frequencies available, the number of arrival times and the telescopes used to provide observations.  We also provide three rankings to indicate which of these pulsars contributes to the clock signals (these final three columns in the Table will be described later). The observational coverage is shown graphically in Figure~\ref{fg:sampling}.  Appendix~\ref{sec:frequentistScript} describes our scripts for processing the data and forming the initial noise model, with Table~\ref{tb:dmModel} listing the noise model parameters.  We note that, as described in the Appendix, some of the PPTA observations were missing from the IPTA data release. These observations were included in our frequentist-based analysis.

With the final parameter and arrival time files, we applied the \citet{hcm+12} technique to derive the common mode, CM($t$), clock signal using a simultaneous global fit to all pulsar data sets\footnote{Note that the term ``common-mode" is being used here in a general manner and refers to the signal common to a group of signals.}.  We constrained the fit for the clock signal so that it does not include an offset, linear or quadratic component.  We also switch off fitting for the pulsar parameters (apart from the pulse frequency and its first time derivative).  The clock, CM($t$), is parameterised by a grid of samples (separated by 0.5\,yr intervals) and linear interpolation between the samples\footnote{We have also re-run the algorithm with a 1\,yr and a 100\,day sampling. These results are available as part of our data release; see Appendix~\ref{sec:dataRelease}.}.

During the work described here, the software {\sc tempo2} (available from \url{https://bitbucket.org/psrsoft/tempo2}) was updated and the default least-squares solver was changed to a scheme using the QR decomposition\footnote{Note that ``QR" is not an acroynm. See, e.g., \cite{ptvf96} for details.} from the previous scheme based on the singular value decomposition (SVD). The QR based solver is faster, but we found a small difference in the clock estimate for the full sample of pulsars. The worst case difference was 0.6 times the error bar. When we used only the eight most significant pulsars the difference between QR and SVD based schemes became insignificant. Further testing revealed that the covariance matrix of the parameters had full rank, but the condition number  of that matrix had increased by a factor of 40 when all pulsars were used. This behaviour is not unexpected. The SVD approach is particularly useful when the covariance matrix of the observations is singular, but it is also  valuable when the whitened model matrix is ill-conditioned. Users of {\sc tempo2} who stress the algorithm with global solutions for many pulsars should take care to test both QR and SVD methods on their problem.  The results presented in this paper were obtained with the SVD approach.

\subsection{Data and processing for the Bayesian analysis}\label{sec:bayesian}

The frequentist method, as described above, relies on the manual modelling of the noise properties, as well as the careful subjective checking of the resulting data sets and fits.  The Bayesian method is independent: it fits both the timing model and the noise models for the pulsars simultaneously.  This avoids the iteration in the frequentist method, however, it requires that the analyst make a choice of prior distributions for the model parameters.

For the Bayesian analysis, we used the IPTA data set without excluding any data from specific observing systems or including the additional PPTA data used in the frequentist method.  We did, however, keep the same time-spans as in the frequentist analysis. Because of the significant computational cost in carrying out the Bayesian analysis, we used a restricted list of eight pulsars.  The selection process is discussed in Section~\ref{sec:whichpsr}. 

Bayesian approaches to analysing pulsar-timing data,
whether to search for spatially correlated signals
or single-pulsar timing and noise analysis, have 
appeared in various publications 
\citep[e.g.][]{vlml09,lah+14,lbj+14}
and have been applied to real data sets  \citep[e.g.][]{dcl+16,cll+16,abb+18}. The results from Bayesian and frequentist methods have been compared and have shown to be generally consistent (for instance, \citealt{lwy+16} and \citealt{bps+16}).

With Bayesian parameter estimation, we estimate the probability distribution of the values
of a set of parameters, $\zeta$, taking
into account the data, $X$, a model (hypothesis), $H$, such as a physical model,
describing how the parameters are related with each other. Bayesian inference is the process by which we use the information from the observed data to  update our knowledge of the probability distribution  of the unknown parameters, for which we necessarily have
a prior belief, described by the prior probability distribution. The inferred distribution is called the posterior probability distribution. Via Bayes' theorem, we perform the parameter estimation using the relation

\begin{equation}
\label{eq:Binfer}
\poster\propto \prior\likeli\,.
\end{equation}

Using $P$ to denote probabilities,  the posterior probability distribution, $\poster=P(\zeta|X,H)$, is therefore the  conditional probability of the parameters  of interest given the model and the data. The likelihood function, $\likeli=\Lambda(X|\zeta,H)$, is the conditional probability  of the data given the model and its parameters. Finally, the prior probability distribution, $\prior{}=P(\zeta|H)$, is the conditional probability of the parameters given only the model hypothesis. The prior is therefore a mathematical tool to express the degree of knowledge or ignorance on the real probability distribution and is mandatory in Bayesian inference. Bayesian inference is often used 
when the model's complexity is high and closed-form solutions are not readily available. In such cases, as is the case with our analysis, the posterior distribution is then inferred by randomly drawing samples from the  prior distribution using Monte-Carlo methods and testing their likelihood. 

For the Bayesian analysis in this work,
the pulsar timing and noise analysis was performed  in the same way as in \cite{cgl+18}. 
The noise model consists of three basic components: the white-noise parameters which regulate the uncertainties of the ToAs,
as well as the red- and DM-noise components.
We summarise the noise model and mathematical details in Appendix~\ref{app:bay}
(see \citealt{cgl+18} for more details).

As in \cite{ltm+15} and \cite{cll+16}, 
we model the clock signal as a stationary, 
power-law, red noise process correlated between the pulsars (a monopolar correlation). The power spectrum of the clock signal is modelled as 
\begin{equation}
\label{eq:specClk}
S_{\clk}(f)=\frac{A_{\clk}^2}{f}{}\left(\frac{f}{f_r}\right)^{2 \alpha_{\clk}}\,,
\end{equation}
where $f$ is the Fourier frequency and $f_r$ is a reference frequency set to 1~yr$^{-1}$. 
In \cite{cll+16} the pulsar noise parameters were held fixed at 
their maximum-likelihood values and the maximum-likelihood values for the clock-noise parameters were calculated in a frequentist 
way, with a maximum-likelihood estimator. 
In this current study, while we still keep the pulsar noise parameters fixed, we perform Bayesian inference for the
clock-error signal parameters while we analytically marginalise over the timing parameters (the analytical marginalisation of the timing parameters is also implicitly assumed in \citealt{cll+16} in the formation of the likelihood function, as in this study). 
Fixing the noise offered the ability to perform the
analysis faster, having first verified that this approach
was sufficient to detect and reconstruct the
simulated TT(TAIx2) signal, as discussed in Section~\ref{sec:results}.

Once we obtain the posterior distributions of the clock-signal parameters, we estimate the waveform and uncertainties of the clock signal with the following procedure:
\begin{itemize}
\item we calculated the 1$\sigma$ (68 per cent credible interval)
boundary of the two-dimensional probability distribution
of the two clock parameters 
%(i.e., its signal amplitude and the spectral index) 
using the posterior distribution obtained from the Bayesian inference.  
\item we constructed the updated total covariance matrix, $\cov_{\textrm{upd}}$, 
by summing the inferred covariance matrix of the clock signal ($\cov_{\clk}$)
and pulsar-noise covariance matrices.  
We used this updated covariance matrix to re-fit the linear timing parameters using generalised least-squares fitting 
in the presence of correlated noise
and calculated the post-fit timing residuals, described by the 
vector $\boldsymbol{t}_{\textrm{gls}}$.
\item we used the method described in \cite{lbj+14} to re-construct clock waveforms from 
for all values of the clock signal's posterior distribution within the 1$\sigma$ boundary.
The clock waveform for each case is then

\begin{equation}
	\tres_{\clk}=\cov_{\clk} \cov_{\textrm{upd}}^{-1} \boldsymbol{t}_{\textrm{gls}}\,.
	\label{eq:waveestimator}
\end{equation}
%%$\mathbf{C}$, $\mathbf{C_{\rm clk}}$ and $\mathbf{t_{gls}}$
The final clock waveform is the average of these waveforms 
and the upper and lower envelope of the waveforms provides the reported 1$\sigma$ boundary.
This is different from the original method, which estimated the maximum-likelihood
waveform and the 1$\sigma$ uncertainty levels as
the standard deviation of the
maximum-likelihood estimator, an approximation that is valid
only if the noise is uncorrelated. Since the noise is in general correlated, this approach gives a more reliable estimate of the clock waveform uncertainties.
\end{itemize}

\section{Results and Discussion}
\label{sec:results}

\begin{figure*}
\includegraphics[angle=-90,width=17cm]{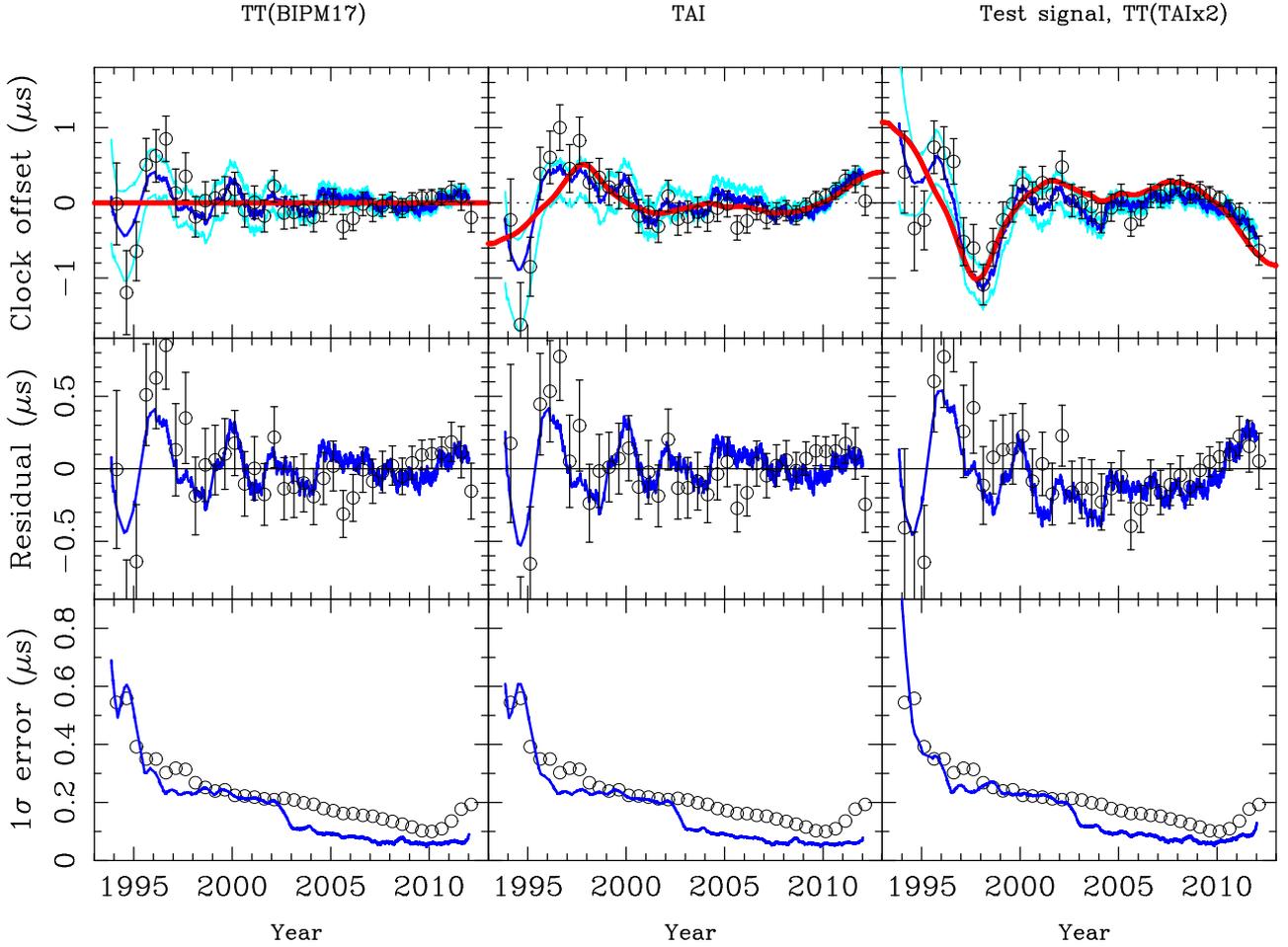}
\caption{The derived clock signal, TT(IPTA16) (top row) referred to TT(BIPM17) (left column), TT(TAI) (central column) and a test reference which is effectively TT(BIPM17)$-2\times$[TT(TAI) - TT(BIPM17)] (right column). The open circles with error bars result from the frequentist analysis and the dark blue lines are the Bayesian determinations of the signal (with 1$\sigma$ uncertainties indicated with the cyan lines). The red line gives the expected signal in each case. The difference between the expected signal and the derived clock signal is shown in the middle row. The uncertainties on the clock signal as a function of time are given in the bottom row for the two analysis methods. }\label{fg:clock}
\end{figure*}

Figure~\ref{fg:clock} shows TT(IPTA16) with respect to three different reference timescales: TT(TAI), TT(BIPM17) and a test case labelled TT(TAIx2).\footnote{This test case was formulated using TT(TAIx2) = TT(TAI)  $-3\times$[TT(TAI) $-$ TT(BIPM17)] as the time reference.}  The three columns correspond to the different reference timescales.   The top row shows the derived clock signal for each reference timescale.  provides the clock waveform.  The black points, with uncertainties, give the frequentist-derived clock signal.  The blue, solid curve indicates the Bayesian waveform with its 1$\sigma$ boundaries in light blue lines.   The expected clock waveform is shown as a solid red line. The second row shows the difference from the expected waveform. The frequentist-derived values are points with errors and the Bayesian values are solid lines.  The third row provides the one-sided error bar, $\sigma$. 

We assume that TT(BIPM17) is an accurate timescale over the sampled data span so the expected clock waveform is zero for time reference TT(BIPM17). For time reference TT(TAI), it is $-$[TT(BIPM17) $-$ TT(TAI)] but with a best-fit quadratic removed. This is necessary because the pulsar timing model must always fit for the spin period and spin-down rate as these are not known a priori. This removes a quadratic polynomial from the residuals of every pulsar, so there can be no quadratic in the resulting clock waveform. This is handled differently in the Bayesian procedure, so we carried out unweighted fits for the quadratic coefficients in both frequentist and Bayesian procedures to make the results comparable. For the third case, the expected signal is $+2\times$[TT(BIPM17) $-$ TT(TAI)] with the quadratic removed.

The error bars on the two analysis schemes are less comparable. The Bayesian lines are 68\% confidence limits on the estimated signal at a given time. The frequentist bars are the estimated standard deviation of that particular sample. They depend on the sampling interval. We also note that the clock signal measurements and uncertainties are not independent. As described by \citet{hcm+12}, the effect of fitting, irregular sampling, differing data spans and the linear interpolation between adjacent grid points all lead to correlated values.  The Bayesian procedure constrains the variation of the clock estimate by its statistical model as a stationary power-law red process, so it is harder to define an interval between independent estimates. The frequentist estimate has a normalized chi-square of 1.08, indicating that the error bars are consistent with the variation in the clock estimate. The Bayesian estimates are also self-consistent.

The following results are easily seen on the figure:
\begin{itemize}
\item the test signals TAI and TAIx2 are recovered within the confidence limits using both frequentist and Bayesian schemes (more details are given below),
\item the residuals (the difference between the expected signal and the derived clock signal) for both methods with TT(BIPM17) and TAI are almost the same, indicating that both algorithms are operating linearly at that signal level -- a much larger signal than expected in TT(BIPM17), 
\item the residuals for TAIx2 are slightly different to those for the preceding two cases, indicating that some non-linearity is becoming evident at this level, i.e., the assumption that the clock signal is small compared to the pulsar noise level starts to fail for this case,
\item the pulsar timescale, TT(IPTA16), with respect to TT(BIPM17) is consistent with zero. Note that the variations of $\pm 500$\,ns before 1998 are not as significant as they appear because the frequentist error bars are 25\% correlated (see Section~\ref{sec:anomalies}). 
\item the IPTA data rate and the receiver performance improved markedly in 2003 and this is reflected in the error bars for both methods, especially in the Bayesian estimates. 
\end{itemize}

Although the known signal in the top central panel of Figure~\ref{fg:clock} is clearly detected, an unknown signal would be less obvious. However an unknown signal of the same variance could be detected purely from the $\chi^2$ of the frequentist samples. The reduced $\chi^2$ of the left panel TT(BIPM17) is 1.08, confirming that the error estimates are good. The reduced $\chi^2$ of the central panel is 2.0 and the number of degrees of freedom is reduced from 37 to 33 by the correlation between samples. Thus a $\chi^2$ of 2.0 indicates a 3.6 $\sigma$ detection of a variance increase.

The mean frequentist error bar size in our clock comparison drops from $\sim$300\,ns in the first half of the data to $\sim$150\,ns in the second half. This implies that our existing data sets could have made a 3$\sigma$ detection of a common signal of $\sim$900\,ns that lasted for 6 months in the early data or $\sim 450$\,ns in the later data.   Longer-lasting events could have been detected at lower amplitudes. For instance, an event lasting 10\,yr with an amplitude $\sim 70$\,ns could be detected in the recent data. We note that large time offsets are not expected from atomic timescales, however frequency instabilities are possible.  At the level that we could detect, such instabilities would likely be caused by planned steering, as in TT(TAI). We have no prior reason to expect any such detectable instabilities in TT(BIPM17).

We have produced a publicly-available data collection containing our input data, processing pipelines and results.  A description of this data collection is given in Appendix~\ref{sec:dataRelease}.

\begin{figure*}
\includegraphics[width=6.5cm,angle=-90]{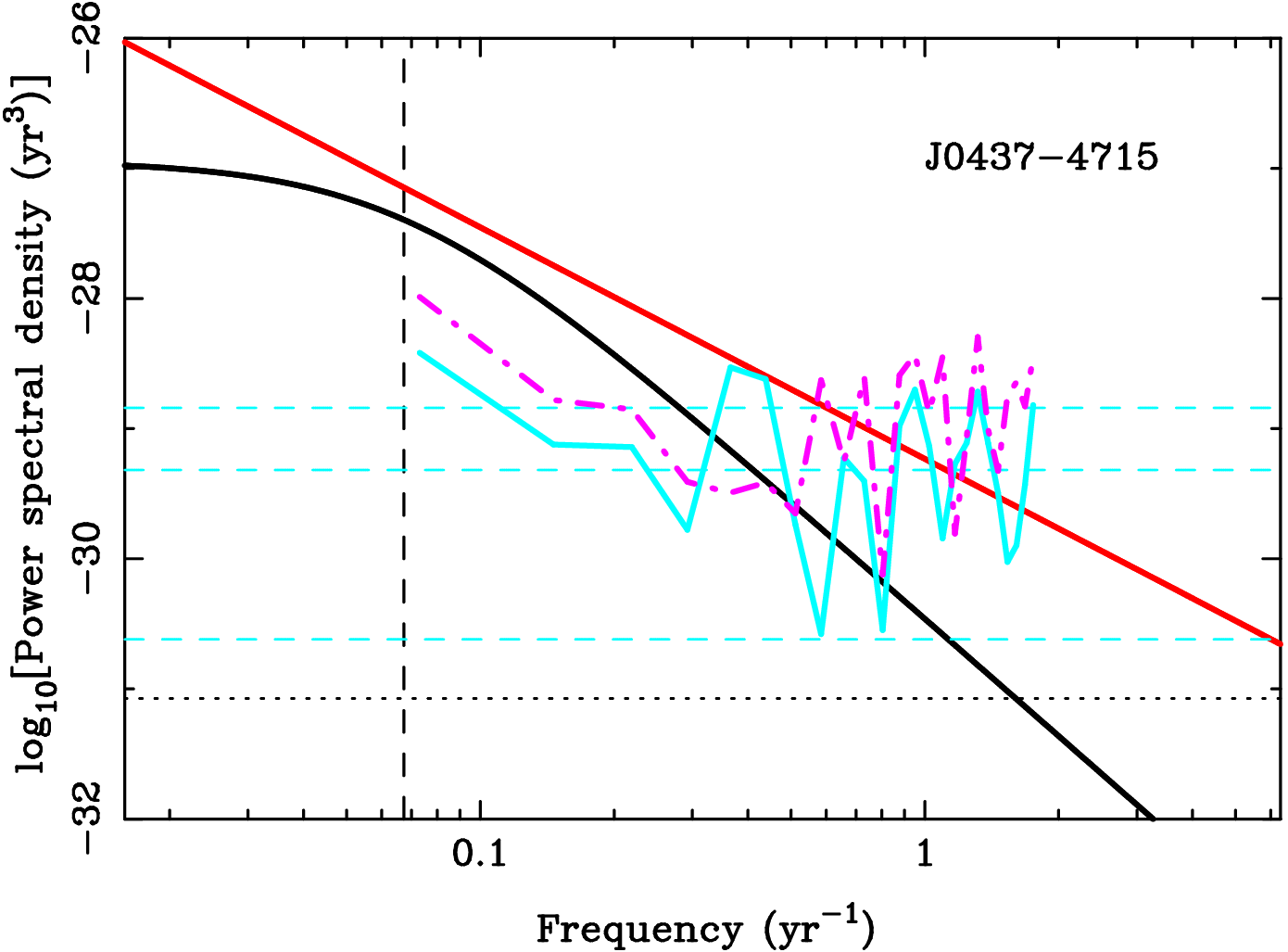}
\includegraphics[width=6.5cm,angle=-90]{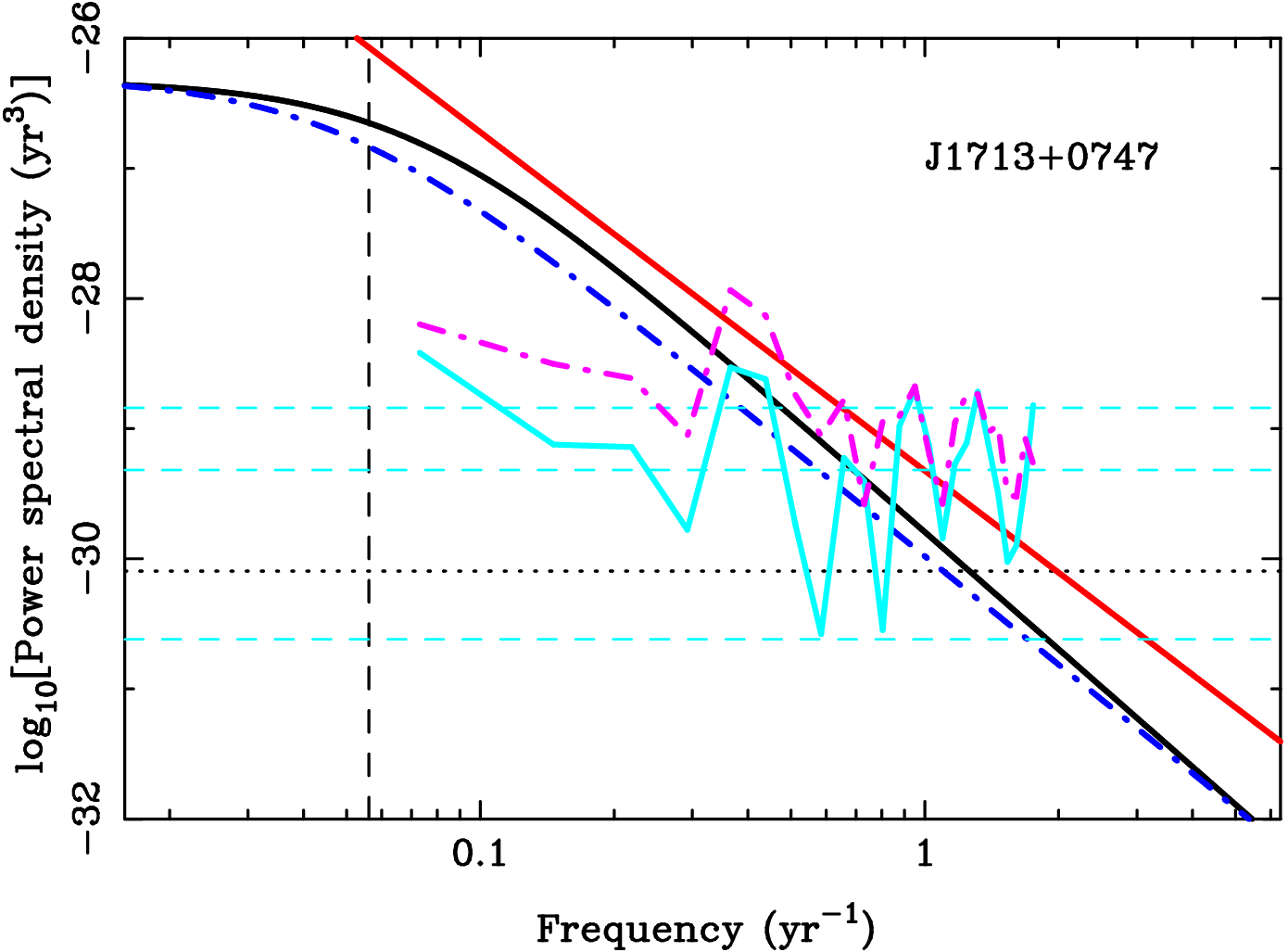}
\includegraphics[width=6.5cm,angle=-90]{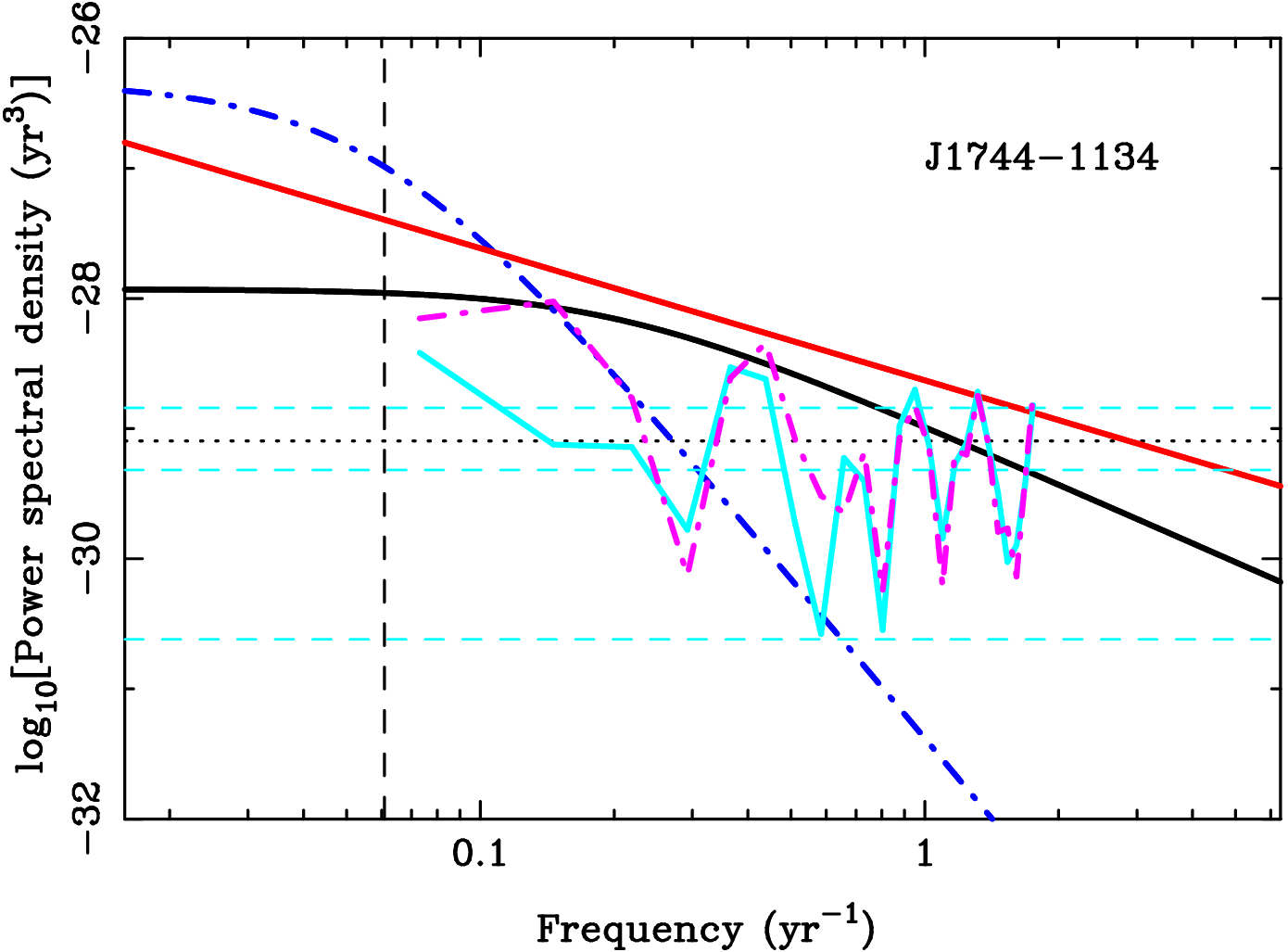}
\includegraphics[width=6.5cm,angle=-90]{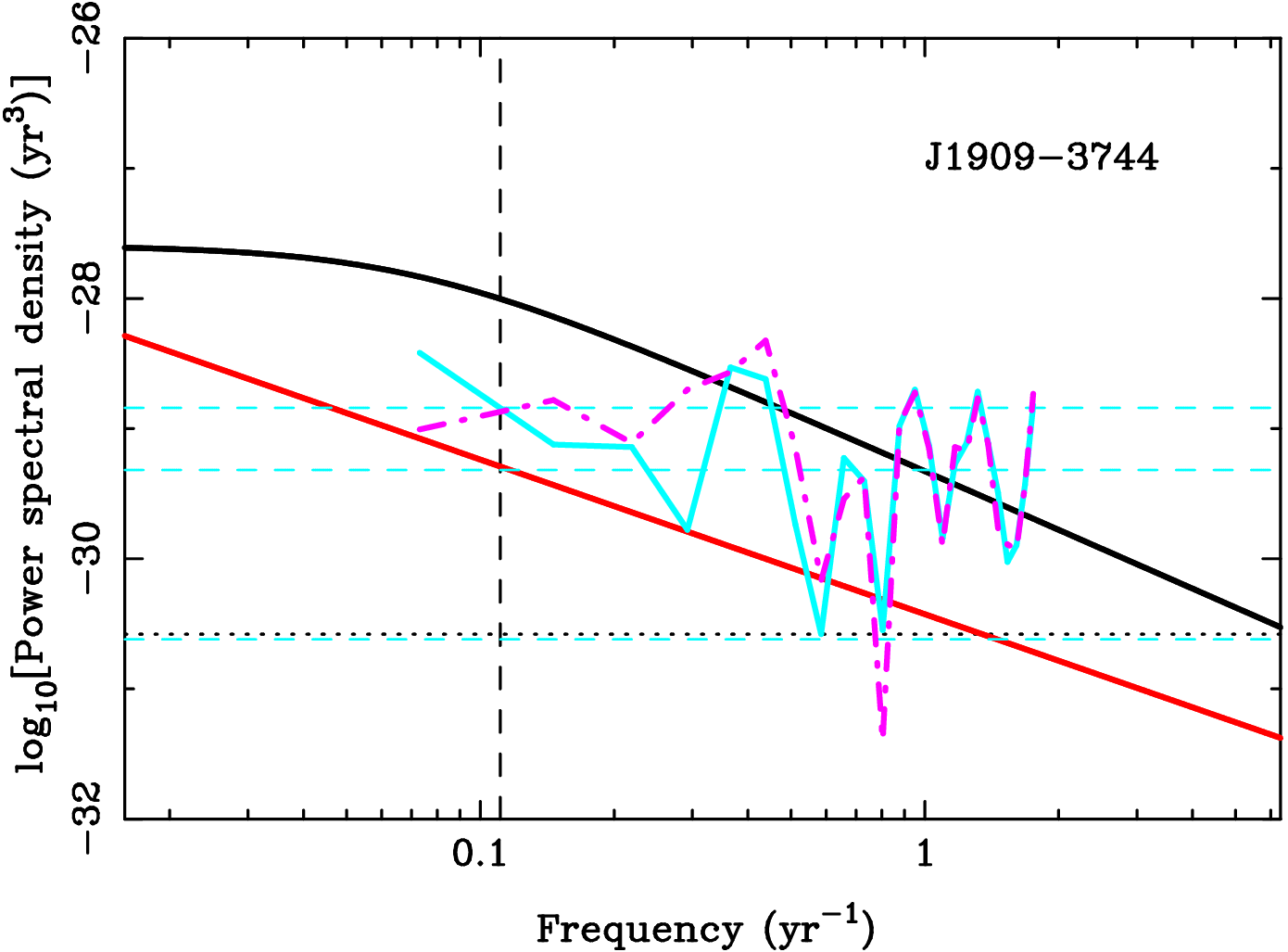}
\caption{Power spectrum of the IPTA clock signal (light blue) with its mean and 95\% confidence levels (horizontal light blue dashed lines) overlaid.  Each panel also contains the red noise models for the specified pulsar from this paper (black for the frequentist analysis and red for the Bayesian analysis) and in the IPTA data release (dot-dashed, blue). The spectral density corresponding to the ToA uncertainties for each pulsar scaled by the white noise error factors is shown as the dotted line.  The purple dot-dashed line indicates the power spectrum of the clock signal after the specified pulsar has been removed from the sample.  The spectral frequency corresponding to $f_0 = 1/[{\rm data\,span}]$ for the specified pulsar is shown as a vertical dashed line.}
\label{fg:clock_cmp}
\end{figure*}

\subsection{Noise power spectra}\label{sec:noiseModel}

The power spectral densities of the noise and the clock estimate are shown in Figure~\ref{fg:clock_cmp}. The blue solid line,  which is the same in each panel of Figure~\ref{fg:clock_cmp}, represents the power spectrum of the frequentist-derived clock (the spectrum of the Bayesian-derived result is shown in Figure~\ref{fg:compareClockEphemGW} and discussed later). The spectrum was formed for data with MJD$> 51000$ (i.e., from the top-left panel in Figure~\ref{fg:clock} ignoring the oldest clock samples; this date corresponds to 6 July 1998).  The horizontal blue dashed lines represent the mean of the spectrum and 95\% confidence levels assuming $\chi_2^2$ statistics (i.e., exponential statistics, with two degrees of freedom).  
Although there is a suggestion that the spectrum might be starting to rise at the lowest frequency sample, the spectrum is consistent with white noise and we will assume it is white in further analyses.

The four panels in this Figure  provide  noise spectra for important pulsars in the sample. The white noise spectrum corresponding to the measured ToA uncertainties that have been corrected for the EFAC and EQUAD parameters (see Equation~\ref{eq:quadratic}) are shown as the horizontal black, dotted line  for the four listed pulsars.   The red noise model as used in the frequentist analysis is shown as the black, solid curve and that from the Bayesian analysis as the red curve.  For comparison we also show the red-noise model for J1713$+$0747 and J1744$-$1134 given in the IPTA data release B (blue dot-dashed curve). The red-noise estimates for PSR~J0437$-$4715 in the IPTA B data set is negligible and no red-noise model was provided for PSR~J1909$-$3744.

The frequentist noise modelling includes a corner frequency in the power spectrum; see Equation~\ref{eq:SpT2}. There was no evidence for a turn-over in the spectrum for PSRs~J0437$-$4715, J1713$+$0747 and J1909$-$3744, but we did not wish to assume a low-frequency variation that we were unable to verify. We therefore used a corner frequency of $f_0 = 1/[{\rm data span}]$. We note that the quadratic fitting procedure applied to all pulsars implies that our choice of noise model below a frequency of $f_0$ has little effect on the final results. 

In contrast, the Bayesian analysis does not include a corner frequency and the Bayesian procedures will estimate signal components with $f < f_0$, albeit with reduced confidence. Comparison between our frequentist-derived noise models and the previous IPTA analysis shows basic consistency, but also that the match is not always close. This highlights the subjective component to the noise analysis.  The Bayesian procedure provides a reproducible method to produce the noise models, but a subjective bias remains in the choice of Bayesian prior.  One key result from our work is that the clock estimates computed with different noise models and different algorithms are consistent within their confidence limits. Clearly the clock fitting process is robust to these changes.

Figure~\ref{fg:clock_cmp} shows that the clock spectrum is much higher than the white noise spectrum for these pulsars below $f = 1$\,yr$^{-1}$. This shows the importance of the red noise models. The clock spectrum for such frequencies is completely dominated by red noise in the pulsar timing residuals.

\subsection{Anomalies}\label{sec:anomalies}

In analysing the data we found occasional anomalies that depended on a few observations of a single pulsar. On checking these carefully we found some errors in the data and either corrected or removed them. However there were a few anomalies remaining for which we could find no reason to remove the data.   One particular example relates to the apparent offset seen in Figure~\ref{fg:clock} around the year 1996. This offset is very similar to that reported by \citet{hcm+12}, which only used PPTA data sets.  

In order to study this anomalously high clock waveform around 1996, we have obtained the frequentist-based clock waveform after removing each pulsar in turn. We find that only PSRs~J0437$-$4715 and J1713$+$0747 significantly affect the pulsar-derived clock signal around this date. If PSR~J1713$+$0747 is removed then the anomaly in the year 1996 disappears (although the uncertainties on the clock signal significantly increase).  Around this time the IPTA data set contains observations from both the Parkes and Effelsberg telescopes and they both indicate a dip of around $2\mu$s.  The observations of PSR~J1713$+$0747 around this time are complex.  The Parkes data were recorded at multiple observing frequencies from 1285 to 1704\,MHz and there is very sparse sampling around the centre of the dip. However, it does seem that an event occurred around this time that is not included in our noise modelling procedures. As we only have a few pulsars contributing to the clock signal at this time, this does lead to an apparent significant clock offset. We know that the timing residuals for PSR~J1713$+$0747 occasionally show unusual behaviour (see, e.g., \citealt{lsc+16}) that cannot be modelled using simple, power-law red noise models.  

The ``bump" in the clock waveform spectrum seen in Figure~\ref{fg:clock_cmp} around frequency 0.4\,yr$^{-1}$ is entirely caused by PSR~J0437$-$4715 (note that the anomaly is missing from the dashed spectrum in Figure~\ref{fg:clock_cmp} produced when PSR~J0437$-$4715 is removed). We have been unsuccessful in determining the cause of this anomaly. We note that this pulsar is only observed by one telescope in the IPTA and so we have no ability, as we do for other pulsars, either to identify and then fix, or to average over, any instrumental effects that may be occurring for a single telescope.

PSR~J1909$-$3744 does not contribute as much as expected in the frequentist analysis, but leads to an anomaly in the clock waveform in the Bayesian analysis around the year 2004.  \citet{srl+15} indicated that the PSR~J1909$-$3744 timing residuals were statistically white, whereas our work indicates red noise.  This is because \citet{srl+15} showed that PPTA observations at high frequencies, which were not corrected for DM variations, produced whiter (and lower rms residual) data sets compared with the DM-corrected lower-frequency data. \cite{lcc+17} also identified excess, observing-frequency-dependent noise in the timing residuals for this pulsar. We therefore believe that uncorrected interstellar medium effects, or instrumental effects are leading to the observed red noise in the IPTA data set for this pulsar.  We repeated the frequentist data processing to re-form the clock waveform after removing the IPTA data-set for PSR~J1909$-$3744 and replacing it with the \citet{srl+15} data set.  Variations of $\lesssim 250$\,ns are seen comparing this clock signal with that shown in Figure~\ref{fg:clock} between the years 2005 and $\sim$2008, which covers the range where this pulsar was observed by the Green Bank Telescope.  The changes that occur when using the IPTA and the \citet{srl+15} data sets and when comparing the Bayesian and frequentist noise estimates highlights both the importance and the challenges of such noise modelling. Such issues have relatively little effect on the pulsar timescale with existing data, but as data sets become longer and ToA precision continues to improve, such modelling will become more important.

\subsection{Which pulsars contribute?}
\label{sec:whichpsr}

The time taken to produce the pulsar-based timescale from both the frequentist and Bayesian algorithms significantly increases as more pulsars are included in the analysis, but not all pulsars are equal.  The pulsars that most constrain the pulsar timescale are those with (1) long data spans, (2) minimal red noise  and (3) few data gaps.

Our next analysis is based on the data shown in the top-left panel of Figure~\ref{fg:clock}. The simplest measure of whether a pulsar contributes is therefore to determine whether the data points in the Figure would change if that pulsar were removed from the analysis.  The maximum change in the frequentist-derived clock signal with the removal of each pulsar is listed in the seventh column  of Table~\ref{tb:summaryTable} along with the ranking of the significance of the pulsar using this measure (from 1 to 48). This ranking highlights that the  pulsars with the largest influence on the resulting clock signal are PSRs~J1713+0747, J1744$-$1134, J0437$-$4715, J2145$-$0750 and J1909$-$3744 each producing variations to the pulsar timescale at a level $>$100\,ns for at least one time sample. PSR~J1713$+$0747 has the largest contribution because the IPTA data release includes the early pre-NANOGrav Green Bank and Arecibo data for this pulsar providing a very long span of high quality data.  There are 22 pulsars whose removal only changes the resulting timescale by $<$10\,ns. Removing all 22 causes no visible change (the maximum change being 13\,ns) to the pulsar-based timescale.  

The analysis described above shows that the removal of some pulsars changes the resulting clock signal.  It does not indicate whether the removal of a pulsar improved the pulsar timescale, or not.  We have considered various statistics to determine whether the inclusion of a pulsar has a positive or detrimental effect.  We have found that comparing the power spectrum of the resulting clock signal (determined on a 100\,d grid) with and without the inclusion of each pulsar in turn indicates most clearly which pulsars are contributing.  

The ratios of the mean of the power spectral densities for $f \leq  0.5$\,yr$^{-1}$ without and with each pulsar in turn are listed in the second last column of Table~\ref{tb:summaryTable}.  The pulsars that are most significant by this measure (in order) PSRs J1713$+$0747, J1744$-$1134, J0437$-$4715, J1022$+$1001, J1909$-$3744 and J0711$-$6830.  In contrast to the first ranking procedure, PSR~J2145$-$0750 is poorly ranked using this method as its inclusion slightly increases the power spectral density of the pulsar-derived clock signal.  We highlight using a bold font (in column 7) those pulsars for which this ratio indicates that the inclusion of the pulsar does not improve the pulsar timescales.

The Bayesian analysis also provided a ranking of pulsars.
This ranking was based on the contribution of each pulsar to the  signal-to-noise ratio of the recovered TT(TAIx2) simulated clock signal using the Bayesian procedure. The procedure was iterative. Starting with only the three pulsars that dominated the initial frequentist analysis (listed as ``Top" in Table~\ref{tb:summaryTable}), we determined the simulated clock signal.
We then repeated the analysis, but each time adding in a different fourth pulsar. We 
chose as the fourth pulsar of the subset the one that increased
the signal-to-noise ratio of the detection the most. We then continued following this method  to add the fifth, sixth, etc. pulsars. The contribution beyond the eighth pulsar was negligible, while adding significant computational time and therefore we
used the eight most dominant pulsars to our analysis. These were the top three, 
PSRs~J0437$-$4715, J1713+0747 and J1909$-$3744, followed by (in order) J1744$-$1134, J1918$-$0642,  J1455$-$3330, J2317+1439 and J1600$-$3053 (the full ranking is given in the last column of  Table~\ref{tb:summaryTable}). The exact choice of pulsars was therefore slightly different in the Bayesian and frequentist methods, but the final results are consistent. 

Both the frequentist and Bayesian tests highlight the importance of 
PSR~J1713$+$0747 as 
the most significant pulsar and then PSRs~J0437$-$4715 and J1744$-$1134 as other key pulsars. This is not a surprise. We have long, high quality data sets on all 
these pulsars. PSR J1909$-$3744 is also important, but it has a shorter data span than the preceding three.

\subsection{Implications for terrestrial time standards}\label{sec:implicartionsTT}

\begin{figure}
    \centering
    \includegraphics[angle=-90,width=8.5cm]{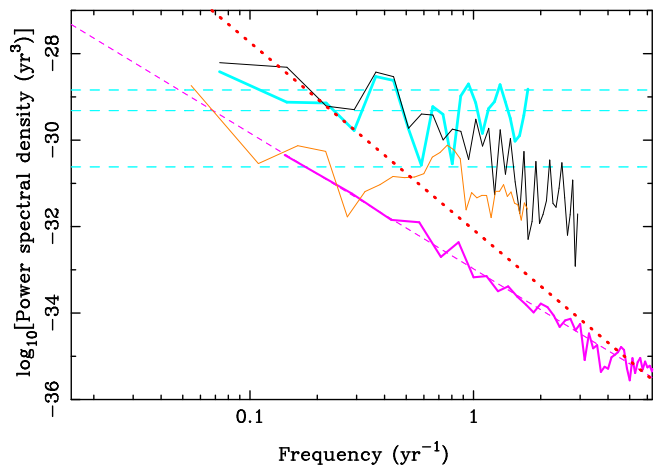}
    \caption{Power spectra of the frequentist-derived (cyan, with mean and 68\% confidence levels indicated  by horizontal dashed lines) and Bayesian-derived (black) pulsar time standard overlaid with the expected power-spectral density of the timing residuals induced by a gravitational wave background with amplitude $10^{-15}$ (red dotted line). Note that only the monopolar contribution of the gravitational wave background signal would affect the measurement of the clock signal. The purple solid line indicates the average power-spectral density of the four USNO rubidium fountain clocks after subtraction of the common-mode signal.  The orange spectrum is derived from the difference in the clock signal when using JPL ephemeris DE436 compared to the default analysis using DE421.}
    \label{fg:compareClockEphemGW}
\end{figure}

Until 2012 the primary clocks used to steer TAI were not operated continuously, so evaluating the stability of the periodically steered time scale required a very sophisticated analysis.  However in 2012 the US Naval Observatory (USNO) began operating a set of four rubidium (Rb) fountains continuously \citep{phs+14} Their time offsets with respect UTC(USNO), which is very close to TAI, were included in the five-day reports which are published monthly by BIPM. We have retrieved these data sets\footnote{The USNO Rb fountains have been submitted to BIPM with the other USNO clocks since 2012. They can be identified by number: 1930002 through 1930005. The data are filed by year and month on the BIPM ftp site, e.g., ftp://62.161.69.5/pub/tai/data/2018/clocks/usno1801.clk corresponds to January 2018.}, which have a 7-year data span, and used them to estimate the stability of the atomic clocks used in generating TAI. The USNO Rb fountains are not `secondary representations of the second', they are operated as clocks, but they are among the most stable clocks in the TAI ensemble \citep{phs+14}. We expect that their stability is at least as good as that of TAI.

The data are uniformly sampled and the noise is stationary so we have used standard spectral analysis to compare the fountain clocks. As the spectral exponents are of order $-$3 we pre-whitened the data with a first difference linear filter to avoid spectral leakage and post-darkened it after the Fourier transformation.  We can eliminate the effects of variations in the reference time scale, any environmental variations common to all of the Rb fountains, and any variations in the time transfer between USNO and BIPM, by comparing the clocks with each other. Rather than do this pairwise, we find a signal common to all four clocks.

We subtract this common-mode signal from each clock independently and, for comparison with the pulsar-derived timescale, we also fit and remove a quadratic before estimating the power spectra. We find that the four Rb fountains corrected for the common-mode signal have similar power spectra, so we took an average of all four to provide the best spectral estimate. The average spectrum, corrected for the common signal, is shown as a solid purple line in Figure~\ref{fg:compareClockEphemGW}. The spectrum of the common-mode signal, which is not shown on the figure, is very similar to that of the average but about a factor of ten higher.

We fitted a power-law model to the corrected average spectrum using a weighted least squares fit, assuming that the spectral estimates were distributed as $\chi_8^2$. The fitted power law model, which has a spectral exponent of $-$3.0, is shown as a dashed purple line. The first sample was not included in the fit because it is partially suppressed by removing the quadratic. Frequencies above 20 y$^{-1}$ were not included in the fit either, as the spectrum begins to flatten at higher frequencies. A spectral exponent of $-$3 is characteristic of `flicker' frequency modulation. The spectrum at frequencies above 20 y$^{-1}$ appears to have an exponent $\sim -$2, which is characteristic of white frequency modulation, but we do not have enough frequency range to establish this clearly.  

%It is unlikely that one clock runs continuously for 15+ years so that extrapolating the stability of one clock is not really useful. Timescales should be used for all long term analyses and ensemble timescales are more stable than individual clocks.
%2. In 10-15 years, as now, a similar analysis may still provide information on the instability of TAI. But such info will already be known by comparison between TAI and TT(BIPM).

The extrapolated mean spectrum of the IPTA pulsar-based timescale (central dashed cyan line) and the extrapolated average spectrum of the Rb fountains intersect at a period of $\sim$15\,yr. However, no individual clock is likely to be operating over this timescale. Consequently, pulsar timing provides a valuable, independent and long-term, check on terrestrial time scales.  The question of whether TT(IPTA) can contribute to the stability of TT(BIPM) is unclear, and probably premature, as it will be more than ten years before its long-term stability can match that of the best Rb fountains -- see, e.g., \citet{ap19} for a recent analysis of the frequency stability of TT(BIPM).

\subsection{The solar system ephemeris and gravitational waves}

The timing residuals processed in this paper were determined using the DE421 solar-system ephemeris (which was also used in the \citealt{vlh+16} publication describing the IPTA data set).  Recent work within the PTAs have highlighted that the choice of solar-system ephemeris has a large effect on the amount of noise seen in the timing residuals \citep[e.g.,][]{abb+18}.  For currently unknown reasons, some of the more recent ephemerides produce noisier pulsar timing residuals than earlier ones for decade-long data sets.   We have reprocessed the frequentist-based analysis using the DE436 solar-system ephemeris\footnote{Available from https://ssd.jpl.nasa.gov/?ephemerides} and find deviations in the resulting clock waveform at the $\pm 50$\,ns level.  The spectrum of the difference between the frequentist-based clock waveforms obtained using DE421 and DE436 is shown as the orange line in Figure~\ref{fg:compareClockEphemGW}. These deviations have no significant effect on the results presented here, but unless the solar-system ephemerides are better understood, they will have a major effect within the next few years.  We note that some of the signatures seen in this spectrum may be related to individual solar system objects and so periodicities may emerge from the spectrum with longer data spans (instead of the low-frequency, red-noise signal, that we currently see).

The primary goal of the IPTA project is to detect ultra-low-frequency gravitational waves (see, e.g., \citealt{btc+18}).  The power spectral density of the timing residuals induced by an isotropic, stochastic, gravitational wave background (GWB) can be approximated as (e.g., \citealt{hjl+09})
\begin{equation}
P(f) = \frac{A^2}{12\pi^2}\left(\frac{f}{f_{\rm 1 yr}}\right)^{-13/3}.
\end{equation}
Current bounds (e.g., \citealt{srl+15}) suggest that $A < 10^{-15}$ and predictions give $A \sim 5 \times 10^{-16}$ \citep{ssbs16}. The present upper bound on the power spectrum of timing residuals that would result from a GWB is shown as a dotted line in Figure ~\ref{fg:compareClockEphemGW}. This spectrum rises above the pulsar clock noise spectrum at a frequency of about 0.2\,yr$^{-1}$ and remains well above the spectrum of the best atomic clocks. This suggests that, unless the GWB amplitude is orders-of-magnitude lower than the current predictions, errors in terrestrial time standards will not be a limiting factor in the detection of gravitational waves.

An unambiguous detection of a GWB requires that the timing residuals for multiple pulsars are shown to be correlated as described by \cite{hd83}. The expected angular correlation has quadrupolar and higher order terms, but its measurement is also affected by (and a GWB affects the measurement of) other spatially correlated processes, including monopolar and dipolar signals (see, e.g., \citealt{thk+16}).  In particular, the expected angular correlation has a nonzero mean of 0.08. This term would therefore appear as a clock error with a spectrum significantly below that of the GWB itself. However if the GWB signal is detected then its effects on the monopole are known and can be subtracted, so the pulsar-based clock can be recovered without distortion by a GWB.

\subsection{Importance of the International Pulsar Timing Array}

The most recent data in the current IPTA data release were obtained in 2012.  The upcoming second data release will have a further five years of data, more pulsars and improved timing precision on the most recent observations.  This new data set should significantly improve the pulsar-based timescale as (1) any extra observations prior to 2005 will help constrain the variations seen in the early data, (2) new, high quality data on a large number of pulsars since 2012 will enable the extension of the timescale and an improved understanding of its long-term stability and (3) up-to-date analysis methods will be applied when producing the data set, which should reduce the number of artefacts and anomalies within the data. Of course, the longer data spans and improved timing precision will also mean that low-frequency noise processes (such as pulsar timing noise) will become more important. 

Millisecond pulsars are known to undergo sudden changes, including small glitch events (\citealt{mjs+16} described a glitch event in the timing residuals of PSR~J0613$-$0200), magnetospheric changes (\citealt{slk+16} reported on such changes for PSR~J1643$-$1224 although note that \citealt{bkm+18} suggest that profile variations in this pulsar may be caused by the interstellar medium and not magnetospheric variations) and effects relating to sudden changes in the interstellar medium (see, e.g., Figure 13 in \citealt{lsc+16} and \citealt{lag+18}).  As our data sets become longer and more extensive, similar events will occur in more pulsars.  Many such events will be challenging to model deterministically and the noise modelling will need to be updated to account better for such non-stationary noise processes. Of course, with sufficient numbers of equally-contributing pulsars, any individual event will be averaged-down, but unless such events can be better understood and modelled they may present a fundamental limit to the stability of the pulsar-based timescale.

In \cite{hcm+12} only observations from the Parkes telescope were used.  This meant that any discrepancies found between the pulsar-based timescale and a terrestrial timescale could arise from errors in the observatory time scale. In contrast the IPTA sample includes data from many observatories, often for a given pulsar.   The most dominant pulsar in the IPTA sample, PSR~J1713$+$0747, includes observations from seven telescopes each with their own independent observatory time scale.  We note that the participating  observatories in the IPTA do not have a well-defined common reference time. For instance, the data processing to form pulse ToAs was carried out using by cross-matching the observations using reference template profiles.  These reference profiles were not identical between the observatories leading to constant phase offsets between results from different observatories.  Similarly, Manchester et al. (2013) describes how derived pulse ToAs from the Parkes observatory are related to the intersection of the azimuth and elevation axes (the topocentric reference point) of the telescope; however, a similar process is not carried out at all the IPTA observatories.  These offsets are accounted for in the fitting process by allowing for an arbitrary phase offset between data from different observatories for each pulsar. Apart from the very earliest data, the observatory-based timescales are transferred to UTC using GPS, and from there to TT(BIPM17). Any inaccuracy in that time transfer would be common to all observatories and any variations that do not take the form of a quadratic-polynomial (and hence, not affected by the fitting procedures) would affect our comparisons between TT(IPTA16) and terrestrial time standards. An analysis in \cite{hcm+12} suggested that any such errors will be $<$10\,ns and therefore have no effect on the results presented here. More detailed studies of the precision for each step in the process from measuring the ToA to the resulting timing residuals have been presented elsewhere (for instance, see the analysis of instrumental noise in \citealt{lsc+16}). We currently see no evidence for a statistically significant common signal in our results (when referred to TT(BIPM17)) and note that it is becoming easier to search for common offsets in PTA data sets as observatory instrumentation becomes more standardised and more telescopes observe the same set of pulsars with similar timing precision.

Pulsar searches are underway at many of the major observatories and we expect the discovery of a large number of new millisecond pulsars.  However, as we have shown in this paper, only a few bright near-by pulsars contribute significantly to the pulsar timescale.  We are not confident that a large number of similarly bright pulsars that can be timed with high precision remain to be discovered. Improving the pulsar time scale by adding more bright pulsars will require use of the next generation of telescopes such as FAST and MeerKAT.

\section{Conclusions}

We have used the IPTA DR1 data set with two different analysis algorithms, one frequentist and one Bayesian, to establish a pulsar-based timescale which we call TT(IPTA16). Consistent results are obtained from the two analysis methods. We show that TT(IPTA16) is  consistent with the timescale TT(BIPM17). We confirm that TT(BIPM17) is currently the most stable timescale in existence and that current efforts to detect nano-Hertz gravitational waves through pulsar timing using a post-corrected BIPM timescale as a reference are not limited by instabilities in the reference timescale.

The pulsar-based time standard will improve with continued
observations using the existing IPTA programs. The IPTA is currently preparing a second data release that includes significantly more pulsars, uses more telescopes and upgraded instruments at the observatories. In the longer term, even more telescopes will start to contribute to the IPTA data set.  Within a few years, we expect high precision observations from the FAST telescope currently being commissioned in China, and the MeerKAT telescope in South Africa.

Of course, atomic clocks will also continue to improve, as fast, or faster, than pulsar observations. However, since pulsar timescales are based on entirely different physics compared to atomic timescales, and are essentially unaffected by any terrestrial phenomena, they form a valuable independent check on atomic timescales. In addition, they provide a timescale that is, in principle, continuous over billions of years.

\section*{Acknowledgements}

The National Radio Astronomy Observatory and Green Bank Observatories are facilities of the NSF operated under cooperative agreement by Associated Universities, Inc. The Arecibo Observatory is operated by SRI International under a cooperative agreement with the NSF (AST-1100968), and in alliance with Ana G. M\'{e}ndez-Universidad Metropolitana, and the Universities Space Research Association.  The Parkes telescope is part of the Australia Telescope which is funded by the Commonwealth Government for operation as a National Facility managed by CSIRO. The Westerbork Synthesis Radio Telescope is operated by the Netherlands Institute for Radio Astronomy (ASTRON) with support from The Netherlands Foundation for Scientific Research NWO. The 100-m Effelsberg Radio Telescope is operated by the
Max-Planck-Institut f\"{u}r Radioastronomie at Effelsberg. Some of the work reported in this paper was supported by the ERC Advanced Grant `LEAP', Grant Agreement Number 227947 (PI Kramer). Pulsar research at the Jodrell Bank Centre for Astrophysics is supported by a consolidated grant from STFC. The Nancay radio telescope is operated by the Paris Observatory, associated with the Centre National de la Recherche Scientifique (CNRS) and acknowledges financial support from the `Programme National de Cosmologie et Galaxies (PNCG)' and `Gravitation, R\'{e}f\'{e}rences, Astronomie, M\'{e}trologie (GRAM)' programmes of CNRS/INSU, France.  LG is supported by  National Natural Science Foundation of China No.11873076. The Flatiron Institute is Supported by the Simons Foundation. The NANOGrav work in this paper was supported by National Science Foundation (NSF) PIRE program award number 0968296 and NSF Physics Frontiers Center award number 1430824. KJL is supported by XDB23010200,
National Basic Research Program of China, 973 Program,
2015CB857101 and NSFC U15311243, 11690024, and the MPG funding for the Max-Planck Partner Group. LG was supported by National Natural Science Foundation of China (NSFC) (nos. U1431117) and \#11010250. EG  acknowledges the support from IMPRS
Bonn/Cologne and the Bonn--Cologne Graduate School
(BCGS). YW acknowledges the support from NSFC under grants 11503007, 91636111 and 11690021. SBS acknowledges the support of NSF EPSCoR award number 1458952. KL and GD acknowledge the financial support by the European Research Council for the ERC Synergy Grant BlackHoleCam under contract no. 610058.
Part of this research was carried out at the Jet Propulsion Laboratory, California Institute of Technology, under a contract with the National Aeronautics and Space Administration.
Work at NRL is supported by NASA.   We thank Patrizia Tavella from the BIPM for helping us understand the nomenclature used when describing time standards.



\bibliographystyle{mnras}
\bibliography{journals,modrefs,psrrefs,crossrefs} % if your bibtex file is called example.bib

\appendix


%%%%%%%%%%%%%%%%%%%%%%%%%%%%%%%%%%%%%%%%%%%%%%%%%%

\section{Frequentist noise modelling}\label{sec:frequentistScript}

Table~\ref{tb:dmModel}  contains, in column order, each pulsar name, the DM, its first time derivative, annual sinusoidal and cosinusoidal components to DM(t), the grid spacing for measuring DM(t), extra DM-covariance parameters ($a$ and $b$; see below) and the band-independent low-frequency noise parameters ($\alpha$, $P_0$ and $f_c$; see below).  

Scripts were used to produce the data products based on the original IPTA data sets. The scripts are available from our data release (see Section~\ref{sec:dataRelease}) and contains \textsc{runProcessScript.tcsh}, which provides the basic processing steps as follows:
\begin{itemize}
\item A check is carried out to ensure that the required software packages are installed.  The frequentist-style processing described here was carried out within a virtual machine on a single laptop.
\item The arrival time and parameter files are corrected for known issues.  Such issues include correcting telescope observing codes, removing any very early or very recent observations (to ensure a well-defined time range for obtaining the clock signal).  IPTA-defined parameters such as phase jumps, white noise models etc. are removed.  The arrival times are subsequently sorted into time order.  PSRs~J0437$-$4715, J0711$-$6830, J1045$-$4509, J1603$-$7202, J1732$-$5049, J1824$-$2452A and J2129$-$5721 are only observed by the PPTA.  For these pulsars we use the more recent \citet{rhc+16} arrival time and parameter files instead of the IPTA data release.  The Parkes observations for PSR~J1909$-$3744 were also taken directly from \cite{rhc+16}.
\item The parameter files were updated to ensure that the JPL DE421 solar ephemeris was used in the timing procedure and the ToAs referred to TT(BIPM17).  Default \textsc{tempo2} parameters (such as models for the dispersion measure variations caused by the solar wind) were used.
\item The NANOGrav data sets are provided in multiple sub-bands.  We have used the \textsc{averageData} plugin in \textsc{tempo2} to produce a single ToA and corresponding uncertainty for each independent observation.  The \textsc{averageData} plugin (1) identifies observations close in time (for us we select all simultaneous observations over different bands observed by NANOGrav), (2) for each block of data the non-weighted mean of the observation times are determined and all the time-dependent parameters in the timing model updates to this epoch, (3) a weighted fit is carried out for a phase offset and its uncertainty using only the specific block of data, (4) a pseudo-observation is added with the arrival time being the new epoch and the frequency and observatory site-code being determined from the  closest, in time, actual observation. The uncertainty of this point is set to the uncertainty on the fitted phase offset, (5) timing residuals are re-formed using this pseudo-data point and the residual subtracted from its arrival time and this process repeated until convergence. A new arrival time file is then produced with only the pseudo-observations. This process has been carefully tested and the long-term timing results (as used when determining the clock signal) are unchanged when the averaged data points are used.
\item The \textsc{efacEquad} plugin to \textsc{tempo2} is run on each observing system for each pulsar.  This plugin determines the white noise corrections ``EFAC" and ``EQUAD" to allow for a scaling error in each ToA uncertainty along with an independent source of white noise. They have the form
\begin{equation}\label{eq:quadratic}
    \sigma^\prime_{i}={\rm EFAC}\ (\sigma_{i}^2+{\rm EQUAD}^2)^{0.5}.
\end{equation}
where $\sigma_i$ is the initial uncertainty for the specified ToA. The residuals for a specific observing instrument and telescope site are ``whitened" using an interpolation scheme (known as ``IFUNCS"). The plugin then trials a well-defined number of EFAC and EQUAD parameters.  For each parameter pair a Kolmogorov-Smirnov (KS) test is carried out to compare the normalised, whitened residuals with Gaussian, white noise. The EFAC/EQUAD parameter pair that yields residuals closest to white, Gaussian noise is subsequently recorded and used for later processing.
\item A new set of arrival time flags are introduced including specific observing bands with bandwidths of 500\,MHz (i.e., 0-500\,MHz, 500-1000\,MHz etc.).
\item We add the data from each observing system into a final arrival-time file in order of data span.  We start with the longest data span and then identify an overlapping data set in the same observing band.  We use the overlapping region to obtain an initial measurement of the offset between the observing systems and then repeat.  
\item Some pulsars have only been observed in a single band.  However, other pulsars have been observed in multiple bands.  For such pulsars we fit for the dispersion measure and its first time derivative as part of the timing model.  Where necessary we also fit for dispersion measure time variations following the \citet{kcs+13} procedure.  
\item Any observations relating to the lowest two bands (0-500\,MHz and 500-1000\,MHz) are then removed as the high-precision timing observations needed for the determining the clock signal are dominated by the higher frequency observations.
\item For the majority of the pulsars we found that a single red noise model (obtained with the \textsc{spectralModel} plugin) was sufficient for modelling the band-independent, low-frequency noise.  The power spectral density, $P(f)$, of the red noise was assumed to have the form
\begin{equation}\label{eq:SpT2}
P(f) = P_0 \ ( 1 + (f / f_c )^2 \ )^{-\alpha/2}.
\end{equation}
An iterative process was subsequently carried out in which the low-frequency noise modelling, jump fitting and parameter estimation were repeatedly checked.   In a few cases, in which the noise was significant, and there was only single-band data in the earlier observations we used the \citet{rhc+16} split-modelling method in which the covariance function of the DM variations were modelled by: 
\begin{equation}\label{eq:KCov}
\rm{Cov}[\rm{DM(\tau)}] = a \exp\left(-(\tau / b\right)^{2}).
\end{equation} For PSRs~J0613$-$0200, J1045$-$4509 and J1643$-$1224 we also modelled annual sinusoidal dispersion measure variations (see \citealt{rhc+16} for details).
\end{itemize}

\begin{table*}
\caption{Parameters describing the DM model (columns 2 -- 8) and the red-noise model (last 3 columns) used for each pulsar. Note that the DM modelling was carried out using all the available observations (including the lowest frequency data, which was removed for subsequent processing).}\label{tb:dmModel}
\begin{tabular}{lcccccccccr}
\hline
Pulsar & DM$_0$ & dDM/dt & Sine  & Cosine & Grid & a & b & $\alpha$ & P$_0$ & f$_c$ \\
 & (cm$^{-3}$pc) & (cm$^{-3}$pc\,yr$^{-1}$) & (cm${-3}$pc) & (cm$^{-3}$pc) & ($\Delta t_{\rm DM}$\,yr) & (s$^2$) & (d) & & (yr$^3$) & (yr$^{-1}$) \\
\hline
J0030+0451 & 4.33  & $3.5\times 10^{-5}$ &  -- &  -- &  -- &  -- & -- & 2  & $6.79851\times 10^{-28}$ &  0.3\\
J0034$-$0534 & 13.77  & $-3.5\times 10^{-5}$ &  -- &  -- &  -- &  -- & -- & 4  & $9.5249\times 10^{-26}$ &  0.5\\
J0218+4232 & 61.25  & $-8.9\times 10^{-4}$ &  -- &  -- &  -- &  -- & -- & 4  & $2.41121\times 10^{-26}$ &  0.6\\
J0437$-$4715 & 2.64  &  -- &  -- &  -- & 0.2  &  -- & -- & 3  & $1.14\times 10^{-27}$ &  0.0673\\
J0610$-$2100 & 60.64  &  -- &  -- &  -- &  -- &  -- & -- & 3  & $3.04161\times 10^{-27}$ &  0.4\\
\\
J0613$-$0200 & 38.78  & $-1.7\times 10^{-4}$ & $-1.0\times 10^{-4}$ & $-1.0\times 10^{-4}$ & 0.3  &  -- & -- & 3  & $9.89339\times 10^{-27}$ &  0.2\\
J0621+1002 & 36.47  & $-1.1\times 10^{-2}$ &  -- &  -- &  -- &  -- & -- & 4  & $3.4345\times 10^{-25}$ &  0.8\\
J0711$-$6830 & 18.41  & $1.0\times 10^{-4}$ &  -- &  -- & 0.5  &  -- & -- & 2.5  & $1.03113\times 10^{-27}$ &  0.6\\
J0751+1807 & 30.25  & $-2.2\times 10^{-4}$ &  -- &  -- &  -- &  -- & -- & 6  & $4.84735\times 10^{-27}$ &  0.6\\
J0900$-$3144 & 75.70  & $-6.2\times 10^{-5}$ &  -- &  -- &  -- &  -- & -- & 2  & $5.0048\times 10^{-27}$ &  0.2\\
\\
J1012+5307 & 9.02  & $3.4\times 10^{-5}$ &  -- &  -- & 0.3  &  -- & -- & 4  & $1.02144\times 10^{-27}$ &  1\\
J1022+1001 & 10.25  & $-4.3\times 10^{-6}$ &  -- &  -- & 0.5  &  -- & -- & 6  & $2.295\times 10^{-27}$ &  1\\
J1024$-$0719 & 6.49  & $1.3\times 10^{-4}$ &  -- &  -- &  -- &  -- & -- & 4  & $6.70908\times 10^{-24}$ &  0.07\\
J1045$-$4509 & 58.14  & $-3.7\times 10^{-3}$ & $-7.3\times 10^{-4}$ & $1.9\times 10^{-4}$ & 0.5  & $1.668\times 10^{-11}$ & 179 & 3  & $2\times 10^{-24}$ &  0.0587\\
J1455$-$3330 & 13.56  & $7.9\times 10^{-4}$ &  -- &  -- &  -- &  -- & -- & 2  & $1.76702\times 10^{-26}$ &  0.2\\
\\
J1600$-$3053 & 52.33  & $-5.9\times 10^{-4}$ &  -- &  -- & 0.3  &  -- & -- & 4  & $1.22878\times 10^{-27}$ &  0.4\\
J1603$-$7202 & 38.05  &  -- &  -- &  -- & 0.3  & $6.276\times 10^{-13}$ & 64 & 2.5  & $1.2\times 10^{-25}$ &  0.065\\
J1640+2224 & 18.42  & $1.5\times 10^{-4}$ &  -- &  -- &  -- &  -- & -- & 4  & $1.42107\times 10^{-27}$ &  1\\
J1643$-$1224 & 62.41  & $-1.2\times 10^{-3}$ & $-2.9\times 10^{-4}$ & $-5.9\times 10^{-4}$ & 1.0  &  -- & -- & 3  & $1.59602\times 10^{-25}$ &  0.1\\
J1713+0747 & 15.99  & $2.7\times 10^{-5}$ &  -- &  -- & 0.3  &  -- & -- & 3  & $4.71009\times 10^{-27}$ &  0.07\\
\\
J1721$-$2457 & 48.62  &  -- &  -- &  -- &  -- &  -- & -- & 4  & $2.4091\times 10^{-25}$ &  1\\
J1730$-$2304 & 9.61  & $3.9\times 10^{-5}$ &  -- &  -- & 0.3  &  -- & -- & 4  & $2.01716\times 10^{-26}$ &  0.3\\
J1732$-$5049 & 56.84  & $8.8\times 10^{-4}$ &  -- &  -- &  -- &  -- & -- & 4  & $1.82694\times 10^{-26}$ &  0.2\\
J1738+0333 & 33.79  &  -- &  -- &  -- &  -- &  -- & -- & 3  & $1.8891\times 10^{-27}$ &  0.5\\
J1744$-$1134 & 3.14  & $-1.7\times 10^{-4}$ &  -- &  -- & 0.3  &  -- & -- & 1.5  & $1.17822\times 10^{-28}$ &  0.2\\
\\
J1751$-$2857 & 42.89  &  -- &  -- &  -- &  -- &  -- & -- & 3  & $1.05861\times 10^{-26}$ &  0.4\\
J1801$-$1417 & 57.21  &  -- &  -- &  -- &  -- &  -- & -- & 4  & $1.33307\times 10^{-25}$ &  0.3\\
J1802$-$2124 & 149.62  & $3.3\times 10^{-4}$ &  -- &  -- &  -- &  -- & -- & 4  & $4.99447\times 10^{-26}$ &  0.4\\
J1804$-$2717 & 24.63  &  -- &  -- &  -- &  -- &  -- & -- & 3  & $1.95346\times 10^{-26}$ &  0.5\\
J1824$-$2452A & 119.89  & $1.1\times 10^{-3}$ &  -- &  -- & 0.5  &  -- & -- & 3.5  & $6.26696\times 10^{-25}$ &  0.17\\
\\
J1843$-$1113 & 59.97  &  -- &  -- &  -- &  -- &  -- & -- & 4  & $2.69402\times 10^{-26}$ &  0.2\\
J1853+1303 & 30.55  &  -- &  -- &  -- &  -- &  -- & -- & 4  & $4.37242\times 10^{-28}$ &  2\\
J1857+0943 & 13.30  & $3.6\times 10^{-4}$ &  -- &  -- &  -- &  -- & -- & 1.5  & $4.26553\times 10^{-27}$ &  0.07\\
J1909$-$3744 & 10.39  & $-3.1\times 10^{-4}$ &  -- &  -- & 0.3  &  -- & -- & 1.5  & $2.55001\times 10^{-28}$ &  0.07\\
J1910+1256 & 38.09  & $3.8\times 10^{-3}$ &  -- &  -- &  -- &  -- & -- & 3  & $5.86889\times 10^{-27}$ &  0.3\\
\\
J1911$-$1114 & 31.07  &  -- &  -- &  -- &  -- &  -- & -- & 3  & $9.95153\times 10^{-28}$ &  0.8\\
J1911+1347 & 30.98  &  -- &  -- &  -- &  -- &  -- & -- & 3  & $3.77936\times 10^{-28}$ &  0.4\\
J1918$-$0642 & 26.61  &  -- &  -- &  -- &  -- &  -- & -- & 4  & $1.71992\times 10^{-27}$ &  0.5\\
J1955+2908 & 104.47  &  -- &  -- &  -- &  -- &  -- & -- & 5  & $9.46096\times 10^{-27}$ &  0.8\\
J2010$-$1323 & 22.18  & $5.5\times 10^{-4}$ &  -- &  -- &  -- &  -- & -- & 4  & $8.09517\times 10^{-28}$ &  0.3\\
\\
J2019+2425 & 17.10  &  -- &  -- &  -- &  -- &  -- & -- & 4  & $3.40917\times 10^{-26}$ &  0.4\\
J2033+1734 & 25.00  &  -- &  -- &  -- &  -- &  -- & -- & 3  & $8.65127\times 10^{-26}$ &  0.4\\
J2124$-$3358 & 4.60  & $9.9\times 10^{-5}$ &  -- &  -- & 0.3  &  -- & -- & 3  & $2.61433\times 10^{-25}$ &  0.07\\
J2129$-$5721 & 31.85  & $-1.6\times 10^{-4}$ &  -- &  -- &  -- &  -- & -- & 3  & $1.3839\times 10^{-27}$ &  0.3\\
J2145$-$0750 & 9.00  & $2.1\times 10^{-4}$ &  -- &  -- & 0.3  &  -- & -- & 3.5  & $2.10773\times 10^{-25}$ &  0.07\\
\\
J2229+2643 & 22.68  & $4.5\times 10^{-4}$ &  -- &  -- &  -- &  -- & -- & 4  & $4.3802\times 10^{-27}$ &  0.7\\
J2317+1439 & 21.90  & $-5.5\times 10^{-4}$ &  -- &  -- & 0.3  &  -- & -- & 3  & $2.04548\times 10^{-24}$ &  0.07\\
J2322+2057 & 13.56  & $5.8\times 10^{-2}$ &  -- &  -- &  -- &  -- & -- & 2  & $1.55902\times 10^{-25}$ &  0.2\\
\hline
\end{tabular}
\end{table*}

\section{Bayesian modelling}
\label{app:bay}

Following the same reasoning as \cite{cgl+18},
we used a simpler noise model for the individual pulsars
than the ones published in \cite{lsc+16}.
Initial timing and noise models are produced
with a joint timing and noise analysis using \mn{} \citep{lah+14},
which utilises the \textsc{tempo2} routines to evaluate the timing model and \cite{fhb09} to perform Bayesian inference of the timing and noise parameters
via a nested-sampling Monte Carlo sampling. 
The analysis and noise model is the same as in \cite{cll+16}.
The uncertainties of the observing systems were weighted using, 
as in the frequentists analysis, a combination of EFAC and EQUAD per observing system
(grouped as in \citealt{lsc+16}). 
Following the same notation as in Eq.~\ref{eq:quadratic},
the rescaled uncertainties for each observing system is

\begin{equation}\label{eq:sigmaBay}
    \sigma^\prime_{i} = \left[(\sigma\cdot{}\textrm{EFAC})^2+\textrm{EQUAD}^2\right]^{0.5}.
\end{equation}
The final noise models were produced using the \ftwo{} package.
The white-noise levels in this stage are then regulated 
by a `global EFAC' per pulsar, a method that is shown to 
perform adequately well \citep{ltm+15}.
The pulsar red noise and stochastic DM variations were modelled 
as stochastic, wide-sense stationary processes, with power-spectra of the form

\begin{equation}
\label{eq:specNoi}
S(f)=\frac{A^2}{f}{}\left(\frac{f}{f_\textrm{c}}\right)^{2 \alpha}\,,
\end{equation}
where the spectrum is fully described by
two parameters, namely the amplitude, $A$, and spectral index, $\alpha$.
These spectra have a sharp cut-off at 1/[dataspan].
For values of spectral indices that are typical for MSPs,
this is sufficient as power at frequencies lower than 1/[dataspan] is
fitted out by timing parameters.
In the red noise case, this is always true due to the
presence of the spin and spin-down parameters \citep{vlml09,lbj+12}.
To make the introduction of such a power-law spectrum in the
model for the stochastic DM-variations, we introduce in our 
timing models for every pulsar a linear and a quadratic temporal variation term
for the DM, that act as analogues to the pulsar spin and spin-down terms \citep{lbj+14}.
The covariance matrices for the stochastic red noise and DM-variation noise 
have elements calculated as \citep{lbj+14}
\begin{equation}
\label{eq:covMr}
C_{\textrm{r},ij} =\int_{1/T}^\infty S_\textrm{r}(f) \cos(2\pi ft_{ij}) \textrm{d}f\ , \textrm{and}
\end{equation}
\begin{equation}
\label{eq:covMdm}
C_{\textrm{dm},ij} =\frac{\kappa^2\int_{1/T}^\infty S_\textrm{d}(f) \cos(2\pi ft_{ij}) \textrm{d}f}{\nu_i^2\nu_j^2}\,.
\end{equation}
In the above equations, the $\textrm{r}$ and $\textrm{dm}$ subscripts denote the case for red noise or DM variations, respectively. The $i$,$j$ indices denote the time epochs and $t_{ij}$ is the time lag between the 
two respective time epochs,  $\nu$ denotes the observing frequency, $f$ is as usual the Fourier frequency,
and $\kappa = 4.15\times10^{-3}$~s. The additional terms in the case of DM variations are due to the 
fact that the ToA delays from the dispersive effects of the interstellar medium are
modelled to follow
the dispersive law of cold homogeneous plasma, %\citep[e.g.][]{ll1960}, 
i.e. the time delay of a signal
at two observing frequencies %, $\nu_1$ and $\nu_2$,
is proportional to the difference of the inverse squares
of those frequencies. %$\nu_{1}^{-2} - \nu_{2}^{-2}$

In the Bayesian determination of the clock signal, 
we reduced the computational cost
by keeping the pulsar noise properties
fixed to the their maximum-likelihood 
values, while performing Monte-Carlo sampling of the clock
parameters using \mn{}.
As detailed in \cite{cll+16}, the likelihood function of the problem 
can be written separating the parameters of stochastic signals of 
interest, i.e., the clock and pulsar noise parameters, and the timing,
deterministic terms  for which we do not need to estimate the 
posterior distributions (nuisance terms). This approach is identical 
to the approach of \cite{vlml09} estimating the parameters of 
stochastic gravitational-wave background. Because the timing 
parameters are linear, we can marginalise over them analytically and 
use the reduced likelihood function \citep{vlml09} as

\begin{equation}
\label{eq:reducedLikFunc2}
%\begin{split}
\Lambda\propto  \frac{1}{\sqrt{|\rm \cov\cov'|}} \times
\exp\left(-\frac{1}{2} \sum_{i,j,I,J} ({ 
	\delta\tres}_{I,i})^{\trs} {C'}_{I,J, i, j} ({\delta\tres}_{J,j})\right)\,,
%\end{split}	
\end{equation}
The $I,J$ indices denote the different pulsars while
the indices $i,j$ denote the different time epochs.
The timing model is now expressed via the vector
$\delta{\tres}=\delta{\tres}_{\textrm{post}} - {\desm}\delta(\boldsymbol{\epsilon})$,
where $\desm{}$ is the timing model's design matrix,
$\delta{\tres}_{\textrm{post}}$ is the post-fit residual vector, 
and $\delta\boldsymbol(\epsilon)$ is a linear perturbation.
The covariance matrix, $\cov$, is the sum of the clock covariance matrix
and the covariance matrices of the pulsar noise components 
(see Appendix~\ref{app:bay} for relevant equations)
and ${\covdash}={\cov}^{-1}-{\cov}^{-1}{\desm}({\desm}^{\trs}{\cov}^{-1}{\desm})^{-1}{\desm}^{\trs}{\cov}^{-1}$.
As a red-noise process, the clock signal will induce on each pulsar an autocorrelation effect
described by $\cov{}_{\clk,ij}$, calculated with an equation analogous to Eq.~\ref{eq:covMr}
for the pulsar red noise. The clock signal is identical for all pulsars
so we can easily add the cross-correlation effect and denote the elements 
of the clock covariance matrix, $\cov_{\clk}$ as
%The clock covariance matrix is
%$C_{\rm clk\,I,J,i,j}=C_{\rm clk}(t_i-t_j)C_{\textrm{clk}\, I,J}$, 
%with C$_{\textrm{clk}\, I,J}=1$ for all $I$,$J$ pairs. 
\begin{equation}
\label{eq:covclk} 
C_{\clk\,I,J,i,j}=C_{\clk,ij}C_{\clk\,I,J}\,\ , \  
%\begin{cases}
C_{\clk\,I,J} = 1 \ \text{,} \forall I \neq J.\\
%C_{\clk\,I,J} = 0 & \text{for } I = J\,.
%\end{cases}
%with C$_{\textrm{clk}\, I,J}=1$ for all $I$,$J$ pairs.
\end{equation}

\section{Releasing our data, processing scripts and results}\label{sec:dataRelease}

The first IPTA data release is available from \url{http://www.ipta4gw.org}.  We have also made available the exact input data used for the frequentist and Bayesian data processing, along with the script used to carry out the frequentist data processing.  Our data release also provides the pulsar-derived clock waveforms and spectra (such as those shown in Figure~\ref{fg:clock_cmp}) for each pulsar.    This data release can be obtained from the IPTA website (\url{http://www.ipta4gw.org}) and also from CSIRO's data archive (\url{https://doi.org/10.25919/5c354f2623ac5}).

\section{Author affiliations}

\emph{\small
$^{1}$ CSIRO Astronomy and Space Science, Australia Telescope National Facility, Box 76, Epping NSW 1710, Australia \\
$^{2}$ Shanghai Astronomical Observatory, CAS, Shanghai 200030, China  \\
$^{3}$ Max-Planck-Institut f{\"u}r Radioastronomie, Auf dem H{\"u}gel 69, D-53121 Bonn, Germany \\
$^{4}$ Kavli institute for Astronomy and Astrophysics, Peking University, Beijing 100871, P.R.China \\
$^{5}$ Department of Electrical and Computer Engineering, University of California at San Diego, La Jolla, CA 92093, USA  \\
$^{6}$ Centre for Astrophysics and Supercomputing, Swinburne University of Technology, PO Box 218, Hawthorn, VIC 3122, Australia \\
$^{7}$ U.S. Naval Observatory, Washington D.C., USA \\
$^{8}$ National Time Service Center, CAS, Xi'an, Shaanxi 710600, China \\
$^{9}$ Astrophysics Science Division, NASA Goddard Space Flight Center, Greenbelt, MD 20771, USA \\
$^{10}$ ASTRON, the Netherlands Institute for Radio Astronomy, Oude Hoogeveensedijk 4, 7991 PD Dwingeloo, the Netherlands \\
$^{11}$ International Centre for Radio Astronomy Research, Curtin University, Bentley, WA 6102, Australia \\
$^{12}$ Cornell Center for Astrophysics and Planetary Science, Cornell University, Ithaca, NY 14853, USA  \\
$^{13}$ Cornell Center for Advanced Computing, Cornell University, Ithaca, NY 14853, USA \\
$^{14}$ Department of Physics and Astronomy, West Virginia University, Morgantown, WV 26506, USA \\
$^{15}$ Center for Gravitational Waves and Cosmology, West Virginia
University, Chestnut Ridge Research Building, Morgantown, WV 26505,
USA \\
$^{16}$ Laboratoire de Physique et Chimie de l'Environnement et de l'Espace LPC2E CNRS-Universit{\'e} d'Orl{\'e}ans, F-45071 Orl{\'e}ans, France \\
$^{17}$ Station de radioastronomie de Nan{\c c}ay, Observatoire de Paris, PSL University, CNRS/INSU F-18330 Nancay, France \\
$^{18}$ Department of Physics, Hillsdale College, 33 E. College Street, Hillsdale, Michigan 49242, USA \\
$^{19}$ University of East Anglia, Norwich Research Park, Norwich NR4 7TJ, United Kingdom\\
$^{20}$ Department of Astrophysics/IMAPP, Radboud University, P.O. Box 9010, 6500 GL Nijmegen, The Netherlands\\
$^{21}$ Jodrell Bank Centre for Astrophysics, University of Manchester, Manchester M13 9PL, UK \\
$^{22}$ Space Science Division, Naval Research Laboratory, Washington, DC 20375-5352, USA \\
$^{23}$ Jet Propulsion Laboratory, California Institute of Technology, Pasadena, California 91109, USA \\
$^{24}$ Green Bank Observatory, P.O. Box 2, Green Bank, WV, 24944, USA \\
$^{25}$ Center for Computational Astrophysics, Flatiron Institute, 162 Fifth Ave, New York, NY 10010, USA \\
$^{26}$ Physics Department, Lafayette College, Easton, PA 18901 USA \\
$^{27}$ Hungarian Academy of Sciences MTA-ELTE ``Extragalatic Astrophysics'' Research Group, Institute of Physics, E\"{o}tv\"{o}s Lor\'{a}nd University, P\'{a}zm\'{a}ny P. s. 1/A, Budapest 1117, Hungary \\
$^{28}$ INAF-Osservatorio Astronomico di Cagliari, via della Scienza 5, I-09047 Selargius (CA), Italy \\
$^{29}$ Università degli Studi di Cagliari, Dip. Fisica, S.P. Monserrato-Sestu Km 0.700, I-09042 Monserrato, Italy \\
$^{30}$ CSIRO Scientific Computing, Australian Technology Park, Locked Bag 9013, Alexandria, NSW 1435, Australia\\
$^{31}$ School of Physics and Astronomy, University of Birmingham, Edgbaston, Birmingham, B15 2TT, UK \\
$^{32}$ Australian Research Council Centre for Excellence for Gravitational-Wave Discovery\\
$^{33}$ Department of Physics and Astronomy, University of British Columbia, 6224 Agricultural Road, Vancouver, BC V6T 1Z1, Canada \\
$^{34}$ Center for Gravitation, Cosmology and Astrophysics, Department of Physics, University of Wisconsin-Milwaukee, PO Box 413, Milwaukee, WI 53201, USA \\
$^{35}$ TAPIR, California Institute of Technology, MC 350-17, Pasadena, CA 91125 USA \\
$^{36}$ Laboratoire Univers et Th{\'e}ories LUTh, Observatoire de Paris, PSL University, CNRS/INSU, Universit{\'e} Paris Diderot, 5 place Jules Janssen, F-92190 Meudon, France  \\
$^{37}$ Xinjiang Astronomical Observatory, Chinese Academy of Sciences, 150 Science 1-Street, Urumqi, Xinjiang 830011, China \\
$^{38}$ School of Physics, Huazhong University of Science and Technology, Wuhan, Hubei Province 430074, China \\
$^{39}$ Monash Centre for Astrophysics (MoCA), School of Physics and Astronomy, Monash University, VIC 3800, Australia \\
}

% Don't change these lines
\bsp	% typesetting comment
\label{lastpage}
\end{document}